\documentclass[11pt]{article}
\usepackage[T1]{fontenc}
\usepackage{graphicx} 

\usepackage[utf8x]{inputenc}
\usepackage{amssymb,amsmath,mathrsfs,enumerate,mathtools}
\usepackage{graphicx,rotate,multicol,braket,multirow}
\usepackage{cite}
\usepackage{braket}
\usepackage{array}
\usepackage{url}
\usepackage{comment}
\usepackage{float}
\usepackage{slashed}
\usepackage[normalem]{ulem}
\usepackage{wrapfig}
\usepackage[textfont=it,labelfont=bf]{caption}
\usepackage{bm}
\usepackage{cancel}
\usepackage[colorlinks=true,
linkcolor=blue,
urlcolor=blue,
citecolor=teal]{hyperref}
\usepackage{lineno}
\usepackage{color}
\usepackage{bbold}
\usepackage{orcidlink}           

\long\def\rpl#1!!#2!!{\textcolor{red}{#1} \textcolor{blue}{#2}}

\def\bar{\overline}
\def\tilde{\widetilde}

\evensidemargin=\oddsidemargin
\title{Probing compressed triplet scalars with ISR jets and soft leptons at the LHC}

\author{
        \sf
        Atri Dey$^{a}$\footnote{atridey1993@gmail.com, atri.dey@physics.uu.se} \ \orcidlink{0000-0002-1645-7641},
        Tathagata Ghosh$^{b,c}$\footnote{tathagataghosh@hri.res.in}\ \orcidlink{0000-0002-8259-0328}, 
        Biswarup Mukhopadhayaya$^{d,e}$\footnote{biswarup@iiserkol.ac.in},
        Agnivo Sarkar$^{b,f}$\footnote{\href{mailto:agnivosarkar@hri.res.in}{agnivosarkar@hri.res.in},\href{mailto:dragnivosarkar@biet.ac.in}{dragnivosarkar@biet.ac.in
}}\ \orcidlink{0000-0001-9596-1936}
       \\[3mm]
       \small\em
       $^a$Department of Physics and Astronomy, Uppsala University, Box 516, SE-751 20 Uppsala, Sweden\\
      \small\em              
        $^b$Regional Centre for Accelerator-based Particle Physics, Harish-Chandra Research Institute,\\
	\small\em 
    Chhatnag Road, Jhunsi, Prayagraj 211019, India\\
    \small\em
    $^c$ Homi Bhabha National Institute, Training School Complex, 
Anushakti Nagar, Mumbai 400 094, India\\
    \small\em
    $^d$School of Physical Sciences, Indian Association for the Cultivation of Science,\\
        \small\em
    2A \& 2B Raja S. C. Mallick Road, Jadavpur, Kolkata - 700032, India \\
    \small\em
         $^e$Department of Physical Sciences,
Indian Institute of Science Education and Research Kolkata, \\
    \small\em 
Mohanpur, 741246, India. \\
        \small\em 
        $^f$Department of Physics, Bharat Institute of Engineering and Technology, Mangalpally, \\
        \small\em 
         Ibrahimpatnam, Ranga Reddy, Telangana 501510, India
        }
\date{}

\begin{document}

\begin{flushright}
	\small{HRI-RECAPP-2026-02}
\end{flushright} 

{\let\newpage\relax\maketitle}

\begin{abstract}
The Type-II seesaw model predicts doubly and singly charged scalars along with neutral Higgs states originating from an $SU(2)$ triplet. Current LHC searches by the ATLAS and CMS collaborations constrain these particles mainly under the assumption that the doubly charged scalar decays dominantly into same-sign dileptons or dibosons. However, when moderate mass splittings exist among the triplet scalars, cascade decays can dominate, suppressing these conventional search channels and leaving sizeable regions of parameter space weakly constrained. We study this compressed region characterized by $1~\text{GeV} \lesssim \Delta M \lesssim 30~\text{GeV}$ and triplet vev $v_t \sim 10^{-7} - 10^{-3}$ GeV. In this scenario, charged scalars predominantly undergo cascade decays, while neutral scalars decay invisibly into neutrinos, leading to final states with soft leptons and missing transverse energy. We propose a dedicated search strategy at the 14 TeV LHC exploiting a hard initial-state radiation jet to boost the scalar system. Using a cut-and-count analysis, we show that discovery-level sensitivity can be achieved in this previously unexplored region with an integrated luminosity of $3000~\mathrm{fb}^{-1}$.
Our results signify the importance of dedicated searches targeting cascade-dominated and compressed mass spectra for beyond the standard model scenarios with an $SU(2)$ multiplet.     

\end{abstract}

\maketitle
\newpage
\section{Introduction}
\label{sec:Intro}

The standard model (SM) of particle physics fails to provide a satisfactory explanation for the observed neutrino masses and mixing. To resolve this shortcoming, the Type-II seesaw mechanism~\cite{Konetschny:1977bn,Cheng:1980qt,Lazarides:1980nt,Schechter:1980gr,Mohapatra:1980yp,Magg:1980ut} offers a well-motivated theoretical framework which is a minimal extension of the SM. In this beyond the standard model (BSM) scenario, the scalar sector is extended with a $SU(2)_{L}$ triplet field ($\Delta$) with hypercharge $Y = 2$ in addition to the usual SM-like scalar doublet ($\Phi$). The doublet field acquires a vacuum expectation value (\emph{vev}) $v_{d}$ at the electroweak scale and breaks the electroweak symmetry spontaneously. As a consequence, the triplet \emph{vev} $v_{t}$ is induced through a trilinear scalar interaction of the form $\Lambda(\Phi^{T}i\sigma_{2}\Delta^{\dagger}\Phi + h.c.)$, where $\Lambda$ is a potential parameter with mass dimension one. After spontaneous symmetry breaking, the scalar spectrum of the model consists of a pair of doubly charged ($H^{\pm\pm}$) and singly charged ($H^{\pm}$) Higgs bosons, along with three neutral Higgs states. Of the neutral scalars, two are CP-even ($H^{0},, h^{0}$), while one is CP-odd ($A^{0}$).
In general, the lightest CP-even scalar state $h^{0}$ is identified with the observed SM-like Higgs boson with mass 125 GeV, which was discovered by the ATLAS and CMS collaborations at the LHC~\cite{ATLAS:2012yve,CMS:2012qbp,ATLAS:2022vkf,CMS:2022dwd}. The gauge symmetry of the model allows the scalar triplet to couple directly with the left-handed lepton doublets. After symmetry breaking, the neutral $CP$-even component of the $\Delta$ field acquires a small but non-zero \emph{vev}, which in turn generates Majorana neutrino masses via the seesaw mechanism. An important phenomenological implication of this BSM scenario is the presence of various exotic scalars, particularly the doubly charged Higgs bosons, which give rise to tantalizing signatures at collider experiments. 



Both ATLAS and CMS collaborations~\cite{ATLAS:2012hi,CMS:2012dun,ATLAS:2014kca,CMS:2014mra,CMS:2016cpz,CMS:2017pet,ATLAS:2017xqs,CMS:2017fhs,ATLAS:2018ceg,ATLAS:2021jol,ATLAS:2022pbd} have performed extensive searches for doubly charged scalars ($H^{\pm\pm}$) which are dominantly produced via different electroweak processes in a model-agnostic manner. Typically, these production mechanisms at the LHC include Drell-Yan pair production as well as single production via Vector Boson fusion (VBF) channels \footnote{The CMS collaboration has searched for the single production of $H^{\pm\pm}$ states that are produced via VBF mode and subsequently decay into same-sign $WW$ final state~\cite{CMS:2014mra}. In this process, both the production and decay vertices are controlled by ${H^{\pm\pm}}{W^{\mp}W^{\mp}}$ coupling. In the case of the Type-II seesaw model, this coupling is proportional to the ratio between the doublet and triplet vev $\frac{v_{t}}{v_{d}}$. In the following, we will describe the present BSM scenario $v_{t} \ll v_{d}$. As a result, this will impose relatively weaker constraints in the model parameter space.}. Due to their distinctive electric charges, the $H^{\pm\pm}$ can generate interesting final states with relatively low SM backgrounds. Following production, the doubly charged scalars can decay either into same-sign dilepton pairs or into a pair of same-sign $W$ bosons, depending on the underlying model parameters. To be specific, for the Type-II seesaw model, the dominant production mode is determined by the triplet \emph{vev} $v_{t}$. In the leptonic decay scenario ($v_{t} \leq 10^{-5}$GeV), assuming an equal branching ratio into all possible same-flavor and mixed-flavor final states, $BR(H^{\pm\pm} \to e^{\pm}e^{\pm}) = BR(H^{\pm\pm} \to \mu^{\pm}e^{\pm}) = BR(H^{\pm\pm} \to e^{\pm}\tau^{\pm}) = BR(H^{\pm\pm} \to \mu^{\pm}\mu^{\pm}) = BR(H^{\pm\pm} \to \mu^{\pm}\tau^{\pm}) = BR(H^{\pm\pm} \to \tau^{\pm}\tau^{\pm})$ the ATLAS collaboration~\cite{ATLAS:2022pbd} has placed a lower bound on the $H^{\pm\pm}$ mass of approximately $m_{H^{\pm\pm}} \lesssim 1080$ GeV. Apart from the leptonic modes, the $H^{\pm\pm}$ can also decay into pair of same sign $W$ boson, $H^{\pm\pm} \to W^{\pm}W^{\pm}$, which becomes particularly dominant for larger value of $v_{t}$ ($v_{t} \geq 10^{-3}$ GeV). For this scenario, dedicated searches exclude $m_{H^{\pm\pm}}$ masses in the range of 200 GeV - 350 GeV~\cite{ATLAS:2021jol} depending upon different modes of $W$-boson decay.  

Several recent studies~\cite{Antusch:2018svb,deMelo:2019asm,Primulando:2019evb,Chun:2019hce,Cai:2017mow,Deppisch:2015qwa} have reinterpreted existing experimental limits to derive the bounds on the parameter space relevant for the scalar sector of the Type-II seesaw model. From their analysis, one can determine the current lower limit on the mass of the doubly charged scalar. As discussed before, these limits primarily depend on the assumption that $H^{\pm\pm}$ decays via a pair of same-sign dilepton or a pair of same-sign $W$ boson modes. Interestingly, lower limits on the $H^{\pm\pm}$ mass can be significantly relaxed if one considers the relative mass gap between the triplet scalars. In this choice, the cascade decay modes among the triplet scalars open up and can appear as dominant channels for a range of $v_{t}$ values. 

To explain the cascade decay mode for the $H^{\pm\pm}$ we like to point out that the scalar sector of this BSM framework is largely governed by three parameters - the mass of the doubly charged scalar $m_{H^{\pm\pm}}$, the mass gap $\Delta M = m_{H^{\pm\pm}} - m_{H^{\pm}} = m_{H^{\pm}} - m_{H/A}$ and the triplet \emph{vev} $v_{t}$. Now for a moderate value of $\Delta M$ (and $v_{t} \sim 10^{-7} - 10^{-3}$ GeV), the $H^{\pm\pm}$ can decays via $H^{\pm\pm} \to H^{\pm}W^{*}$. In Ref.~\cite{Ashanujjaman:2021txz}, the authors have shown that for a moderate mass gap, which is delineated as 1 GeV $\lesssim \Delta M \lesssim 30$ GeV, the charged triplet scalars can predominantly decay via cascade modes. As a consequence, the aforementioned conventional channels that are analysed in different direct search experiments have suppressed branching ratios. In addition, for $v_{t} \leq 10^{-4}$ GeV, both the neutral BSM scalars can decay via the $\bar{\nu}^{c}\nu$ mode. As a result, a non-trivial region of parameter space, which is described as $\Delta M \gtrsim 10$ GeV and $v_{t} \sim 10^{-7} - 10^{-3}$ GeV, remains effectively unconstrained by the existing limits.

In this work, we target this largely unexplored region of the Type-II seesaw parameter space, where conventional search strategies lose sensitivity due to the compressed scalar spectrum and the presence of soft decay products. We re-iterate that the occurrence of this region is the result of an interplay of kinematics and dynamics, consistently with the quest for an explanation of neutrino masses via the type-II seesaw mechanism. We develop a dedicated collider search strategy tailored to this regime, demonstrating that a focused and physically motivated simple cut-and-count analysis can effectively probe this challenging parameter space.

Probing this region is challenging for several reasons, which we outline below.
\begin{itemize}
     
    \item The widths for both $H^{\pm\pm} \to \ell^{\pm} \ell^{\pm}$ and $H^{\pm\pm} \to W^{\pm}W^{\pm}$ decays are relatively small. Consequently, $H^{\pm\pm}$ predominantly decays via the channel $H^{\pm\pm} \to  W^{{\pm}^*} H^{\pm}$. Similarly, the singly charged scalar dominantly decays through $H^{\pm} \to W^{{\pm}^*} H^{0}/A^{0}$. Finally, in the aforementioned unconstrained region, all neutral triplet scalars decay exclusively via the $H^{0}/A^{0} \to \bar{\nu}^{c}\nu$ mode.
    
    \item The triplet scalar mass spectrum is compressed. As a result, the $W^{\pm}$ bosons produced in the cascade decays of charged scalars are kinematically off-shell. Consequently, the visible daughter jets and leptons arising from subsequent decays are very soft and fail to pass the LHC thresholds. Recently, Ref.~\cite{Ashanujjaman:2023tlj} presented an analysis targeting this parameter space. In that work, the authors focused on same-sign dilepton final states. However, we find the dilepton trigger employed there to be unrealistic for the LHC. Triggering such soft events at the LHC without recoiling against a hard object remains a significant challenge.

    \item In addition, the heavy charged scalars are produced back-to-back, and the missing energy, $E^{\text{miss}}_{T}$, generated from their decay largely cancels out, thereby limiting the usefulness of a large $E^{\text{miss}}_{T}$ as a trigger.     
\end{itemize}

In our work, we explore the prospects of probing $SU(2)$ triplet scalars with compressed spectra in association with a hard initial state radiation (ISR) jet. The recoil against a hard QCD jet induces an opening angle between the pair of charged scalars, leading to a significant enhancement in $E^{\text{miss}}_{T}$ in the system. Consequently, this setup enables the use of an $E^{\text{miss}}_{T}$ trigger~\cite{ATLAS:2016wtr}, which has proven effective in probing compressed mass spectra scenarios in supersymmetric models~\cite{ATLAS:2021moa}. To balance the large $p_T$ of the ISR jet, the triplet scalars acquire a boost, rendering the leptons from their decays sufficiently energetic to pass LHC thresholds in conjunction with the $E^{\text{miss}}_{T}$ trigger. In summary, our approach exploits an ISR jet in association with missing transverse energy and soft leptons, thereby significantly enhancing the accessibility of compressed spectra scenarios. We demonstrate that this strategy can extend the LHC sensitivity well beyond the canonical parameter space explored so far.


This paper is organized as follows. In Sec.~(\ref{sec:setup}), we briefly describe the scalar sector of the Type-II seesaw model. After that, we determine that the benchmark scenarios are safe from existing experiments and present them in Sec.~(\ref{sec:conbench}). Furthermore, in Sec.~(\ref{sec:cut&count}), we present our analysis strategy, including the definition of selection cuts and a cut-and-count analysis, and discuss the resulting signal significance. Finally, we summarize our findings in Sec.~(\ref{sec:concl}).


\section{A Brief Description of the Seesaw Scalar Sector}
\label{sec:setup}

In this section, we present the essential aspects of the Type-II seesaw model with particular attention to its scalar sector. In addition to the SM-like scalar doublet $\Phi$, this model contains an additional $SU(2)_{L}$ scalar triplet field $\Delta$ with hypercharge $Y = 2$. In gauge basis, the triplet and doublet fields can be written in the following manner:  
\begin{equation}
\Delta = \begin{bmatrix}
\delta^{+}/\sqrt{2} & \delta^{++} \\
\delta^{0} & - \delta^{+}/\sqrt{2}
\end{bmatrix} ~~~~~~~ \text{and} ~~~~~~  \Phi = \begin{bmatrix}
    \phi^{+} \\
    \phi^{0}
\end{bmatrix}.
\label{Eq:tripletform}
\end{equation}

In Eq.~(\ref{Eq:Potential}) we write down the gauge invariant scalar potential following the notation described in Ref.~\cite{BhupalDev:2013xol,Dev:2017ouk}. 
\begin{equation}
\begin{split}
    V(\Phi, \Delta) & = -\mu^{2}\Phi^{\dagger}\Phi + \frac{\lambda}{2}\left(\Phi^{\dagger}\Phi\right)^2 +\tilde{M}^{2}_{\Delta}\text{Tr}\left(\Delta^{\dagger}\Delta\right) + \frac{\lambda_{1}}{2}\text{Tr}\left[\left(\Delta^{\dagger}\Delta\right)\right]^{2} \\
    & + \frac{\lambda_{2}}{2}\left(\text{Tr}\left[\left(\Delta^{\dagger}\Delta\right)\right]^{2} - \text{Tr}\left[\left(\Delta^{\dagger}\Delta\right)^{2}\right]\right) + \lambda_{4}\left(\Phi^{\dagger}\Phi\right)\text{Tr}\left(\Delta^{\dagger}\Delta\right) + \lambda_{5}\Phi^{\dagger}\left[\Delta^{\dagger}, \Delta\right]\Phi  \\
   & + \left(\frac{\Lambda_{6}i}{\sqrt{2}}\Phi^{T}\sigma_{2}\Delta^{\dagger}\Phi + \text{h.c}\right). 
\end{split}
\label{Eq:Potential}
\end{equation}
Here, $\mu^{2}$, $\tilde{M}^{2}_{\Delta}$, and $\Lambda_{6}$ denote the dimension-full mass parameters appearing in the scalar potential. The quartic couplings $\lambda$ and $\lambda_{i}$ are independent dimensionless parameters and, without loss of generality, can be considered to be real by assuming an appropriate phase redefinition of the triplet field $\Delta$. 
The neutral component of the $\Phi$ field acquires a non-zero \emph{vev} $v_d$, thereby triggering electroweak symmetry breaking. The trilinear interaction term involving $\Lambda_6$ in Eq.~(\ref{Eq:Potential}) induces a tadpole term for the triplet field, which subsequently generates a non-zero \emph{vev} $v_t$ for the neutral component of $\Delta$.
Minimising the scalar potential with respect to the neutral scalar fields leads to two independent tadpole conditions, as given below 
\begin{align}
    \mu^{2} &= \frac{\lambda}{2}v^{2}_{d} - \Lambda_{6}v_{t} + \frac{1}{2}\left(\lambda_{4} - \lambda_{5}\right)v^{2}_{t} \nonumber \\
    \tilde{M}^{2}_{\Delta} & = \frac{\Lambda_{6}v^{2}_{d}}{2v_{t}} - \frac{\lambda_{1}}{2}v^{2}_{t} - \frac{1}{2}\left(\lambda_{4} - \lambda_{5}\right)v^{2}_{d} \, .
    \label{Eq:tadpole}
\end{align}
These identities relate the dimension-full scalar parameters to the doublet \emph{vev} $v_{d}$ and triplet \emph{vev} $v_{t}$ respectively.

These conditions will be useful to determine the mass eigenvalue of different scalar fields. The gauge interaction of these scalar fields is governed by their corresponding covariant derivatives -
\begin{align}
    D_{\mu}\Phi &= \left(\partial_{\mu} -i\frac{g}{2}\sigma^{a}W^{a}_{\mu} - i\frac{g^{'}}{2}B_{\mu}\right)\Phi, \nonumber \\
    D_{\mu}\Delta &= \partial_{\mu}\Delta - i\frac{g}{2}[\sigma^{a}W^{a}_{\mu}, \Delta] - i\frac{g^{'}}{2}B_{\mu}\Delta,
    \label{Eq:cov}
\end{align}
where $g$, $g^{'}$ are the $SU(2)_{L}$ and $U(1)_{Y}$ gauge couplings respectively. After electroweak symmetry breaking, the kinetic terms $(D_{\mu}\Phi)^{\dagger}(D^{\mu}\Phi)$ and $\text{Tr}[(D_{\mu}\Delta)^{\dagger}(D^{\mu}\Delta)]$ generate the mass terms for the electroweak gauge bosons with both $v_{d}$ and $v_{t}$ are contributing to the $W$ and $Z$ boson masses. At the tree level, these masses can be expressed as 
\begin{equation}
    M^{2}_{W} = \frac{g^{2}}{4}(v^{2}_{d} + 2v^{2}_{t}) \, , ~~~~ M^{2}_{Z} = \frac{g^{2} + g^{'2}}{2}(v^{2}_{d} + 4v^{2}_{t}) \, .
    \label{Eq:wzmass}
\end{equation}
The corresponding tree level $\rho$ parameter in this model takes the following form 
\begin{equation}
    \rho \equiv \frac{M^{2}_{W}}{ M^{2}_{Z}\cos^{2}\theta_{W}} = \frac{v^{2}_{d} + 2v^{2}_{t}}{v^{2}_{d} + 4v^{2}_{t}}.
\end{equation}
From precision electroweak data, the measured value of the $\rho$ parameter is $\rho = 1.00038(20)$ \cite{ParticleDataGroup:2020ssz}. To satisfy this, the tree-level value of the $\rho$ parameter has to be very close to unity. Using this fact, one can obtain the upper limit on the triplet \emph{vev} $v_{t} \lesssim 0.78$ GeV at $2 \sigma$. Consequently, we work in the phenomenologically viable limit $v_{t} \ll v_{d}$ and identify $v_{d} \to v$, where $v \simeq 246$ GeV denotes the electroweak \emph{vev}. In this limit, the tadpole Eq.~(\ref{Eq:tadpole}) can further be simplified and re-expressed in the following manner -  
\begin{equation}
    v_{t} = \frac{\Lambda_{6}v^{2}}{2\tilde{M}^{2}_{\Delta} + v^{2}\left(\lambda_{4} - \lambda_{5}\right)} \, .
\end{equation}

The gauge symmetry of the model allows us to write down the Yukawa interaction term between the scalar triplet $\Delta$ and the SM lepton doublets $L = (\nu_{L}, \ell_{L})^{T}$ in the following manner:  
\begin{equation}
    \mathcal{L}_{\nu} = Y^{\nu}_{ij}L^{T}_{i}Ci\sigma_{2}\Delta L_{j} + h.c. \, .
    \label{Eq:neutrinomass}
\end{equation}
Here $Y^{\nu}$ is the complex symmetric matrix, $i$ and $j$ are the flavor indices, and $C$ denotes the charge conjugation operator. One can realise that the triplet \emph{vev} $v_{t}$ can generate Majorana mass terms for the neutrinos. The resulting neutrino mass matrix can be expressed as $m^{\nu}_{ij} = \sqrt{2}Y^{\nu}_{ij}v_{t}$. This mass matrix can be diagonalized using the Pontecorvo–Maki–Nakagawa–Sakata (PMNS) matrix ( $U_{PMNS}$), which encodes the observed neutrino mixing angles and CP-violating phases. Explicitly, $U_{PMNS}$ contains three mixing angles $(\theta_{12}, \theta_{13}, \theta_{23})$, on Dirac CP phase $\delta_{CP}$ and two Majorana phase ($\alpha_{1}, \alpha_{2}$) such that $m^{\nu}_{ij} = U_{PMNS}m^{diag}_{\nu}U^{T}_{PMNS}$. Given a specific triplet vev $v_{t}$ and the low-energy neutrino data, this relation allows one to reconstruct the structure of the Yukawa coupling matrix $Y_{\nu}$. For the present study, we adopt the best-fit values of the neutrino oscillation parameters from Ref.~\cite{Esteban:2020cvm}.   

After electroweak symmetry breaking, the neutral scalar states $\phi^{0}$ and $\delta^{0}$ mix to form two CP-even states, $h^{0}$ and $H^{0}$ and two CP-odd state $G^{0}$ and $A^{0}$ with the mixing angle $\alpha$ and $\beta_{0}$ respectively. In the following, we identify the lighter CP-even state $h^{0}$ as the observed SM-like Higgs boson. Similarly, the singly charged states $\phi^{\pm}$ and $\delta^{\pm}$ mix into the mass eigenstates $G^{\pm}$ and $H^{\pm}$ with a mixing angle $\beta_{\pm}$. In contrast to that, the doubly charged scalar gauge states $\delta^{\pm\pm}$ are already aligned with the mass eigenstates $H^{\pm\pm}$. Hereafter, we use $H^{\pm\pm}$ to denote the doubly charged Higgs boson. The Goldstone modes $G^{0}$ and $G^{\pm}$ are absorbed by the $Z$ and $W$ bosons, respectively, providing their longitudinal components and thereby giving them mass. The relationship between the gauge eigenstates and the corresponding mass eigenstates can be expressed as:  
\begin{align}
    H^{\pm\pm} &= \delta^{\pm\pm} \nonumber \\
    \begin{bmatrix}
        G^{\pm} \\
        H^{\pm}
    \end{bmatrix} & = 
    \begin{bmatrix}
        \cos\beta_{\pm} & \sin\beta_{\pm} \\
        -\sin\beta_{\pm} & \cos\beta_{\pm}
    \end{bmatrix} \begin{bmatrix}
        \phi^{\pm} \\
        \delta^{\pm}
    \end{bmatrix} \nonumber \\
     \begin{bmatrix}
        G^{0} \\
        A^{0}
    \end{bmatrix} & = 
    \begin{bmatrix}
        \cos\beta_{0} & \sin\beta_{0} \\
        -\sin\beta_{0} & \cos\beta_{0}
    \end{bmatrix} \begin{bmatrix}
        \text{Im}[\phi^{0}] \\
        \text{Im}[\delta^{0}]
    \end{bmatrix} \\
     \begin{bmatrix}
        h^{0} \\
        H^{0}
    \end{bmatrix} & = 
    \begin{bmatrix}
        \cos\alpha & \sin\alpha \\
        -\sin\alpha & \cos\alpha
    \end{bmatrix} \begin{bmatrix}
        \text{Re}[\phi^{0}] \\
        \text{Re}[\delta^{0}]
    \end{bmatrix} \nonumber \, ,
    \label{Eq:mixing}
\end{align}
where the mixing angles are defined as
\begin{equation}
  \tan\beta_{0} = \sqrt{2}\tan\beta_{\pm} = \frac{2v_{t}}{v_{d}} ~~~~ \text{and} ~~~~ \tan2\alpha = \frac{4v_{t}}{v_{d}}\frac{(\tilde{M}^{2}_{\Delta} + \frac{1}{2}\lambda_{1}v^{2}_{t})}{\tilde{M}^{2}_{\Delta} + \frac{3}{2}\lambda_{1}v^{2}_{t} + \frac{1}{2}(\lambda_{4} - \lambda_{5} - 2\lambda)v^{2}_{t}}   \, .
\end{equation}

In the limit $v_{t} \ll v_{d}$, one of the mixing angles and the masses of the BSM scalars\footnote{One can realise that in this limit, both $\tan\beta_{0}$ and $\tan\beta_{\pm}$ are very small. Here we fix the corresponding values at $\mathcal{O}(10^{-8})$ and light $CP$-even scalar mass becomes $m^{2}_{h^{0}} \simeq 2\lambda v^{2}_{d}$.} would take the following simplified form -  
\begin{align}
    m^{2}_{H^{\pm\pm}} & \simeq \tilde{M}^{2}_{\Delta} + \frac{1}{2}\left(\lambda_{4} + \lambda_{5}\right)v^{2}_{d} \nonumber \\
    m^{^2}_{H^{\pm}} & \simeq \tilde{M}^{2}_{\Delta} + \frac{1}{2}\lambda_{4}v^{2}_{d}     
\nonumber \\
    m^{2}_{A^{0}/H^{0}} & \simeq \tilde{M}^{2}_{\Delta} + \frac{1}{2}\left(\lambda_{4} - \lambda_{5}\right)v^{2}_{d}   \label{Eq:masssvtvd} \\
    \tan2\alpha & \simeq \frac{4v_{t}}{v_{d}}\left(1 - \frac{m^{2}_{h^{0}}}{m^{2}_{H^{0}}}\right)^{-1} \nonumber \\
    M^{2}_{\Delta} & = \tilde{M}^{2}_{\Delta} + \frac{\lambda_{4}}{2}v^{2}_{d} \nonumber  \, .
\end{align}
From Eq.~(\ref{Eq:masssvtvd}), one can realise that the mass-squared difference between these BSM scalars can be expressed as -
\begin{equation}
    m^{2}_{H^{\pm\pm}} - m^{2}_{H^{\pm}} = m^{2}_{H^{\pm}} - m^{2}_{H^{0}/A^{0}} = \frac{\lambda_{5}}{2}v^{2}_{d} \, .
    \label{Eq:massdiff}
\end{equation}
For our later discussion, we would introduce a new parameter which encodes the mass splitting between these BSM scalars, $\Delta M = m_{H^{\pm\pm}} - m_{H^{\pm}} = m_{H^{\pm}} - m_{H^{0}/A^{0}}$. 

Before concluding this section, we would like to mention the theoretical constraints on these scalar parameters. Demanding perturbative unitarity and stability, one can determine these theoretical constraints on the quartic couplings of the scalar potential. The necessary and
sufficient conditions for the potential to be bounded-from-below (BFB) Ref.~\cite{Dev:2017ouk} are given as   
\begin{eqnarray}
        \lambda \geq 0 \,, \quad \lambda_{1} \geq 0 \,, \quad 2\lambda_{1} + \lambda_{2} \geq 0 \,, \quad \lambda_{4} \pm \lambda_{5} + \sqrt{\lambda\lambda_{1}} \geq 0 \,, \quad 2|\lambda_{5}|\sqrt{\lambda_{1}} + \lambda_{2}\sqrt{\lambda} \geq 0 \,, \nonumber\\
        \lambda_{4} + \sqrt{(\lambda\lambda_{2} + 2\lambda^{2}_{5})(\lambda_{1}/\lambda_{2} + 1/2)} \geq 0 \, , \quad \lambda_{4} \pm \lambda_{5} + \sqrt{\lambda(\lambda_{1} + \frac{\lambda_{2}}{2})} \geq 0 \, .
\label{eq:stability}
\end{eqnarray}

Along with that, another set of constraints on scalar potential parameters can be determined by demanding tree-level unitarity to be preserved, while considering different tree-level $2 \to 2$ scattering processes: scalar-scalar scatterings, gauge boson scatterings, and scalar-gauge boson scatterings. For the Type-II seesaw model, to ensure perturbative unitarity.  scalar parameters must obey the following conditions~\cite{Dev:2017ouk,Das:2016bir}
\begin{eqnarray}
\lambda \leq \frac{8}{3}\pi \,, \quad \lambda_{1} - \lambda_{2} \leq 8\pi \,, \quad 4\lambda_{1} + \lambda_{2} \leq 8\pi \,, \quad 2\lambda_{1} + 3\lambda_{2} \leq 16\pi \, , \nonumber  \\ 
|\lambda_{5}| \leq \frac{1}{2}\text{min}\left[\sqrt{(\lambda \pm 8\pi)(\lambda_{1} - \lambda_{2} \pm 8\pi)}\right]  \, , \quad
|\lambda_{4}| \leq \frac{1}{\sqrt{2}}\sqrt{(\lambda - \frac{8}{3}\pi)(4\lambda_{1} + \lambda_{2} - 8\pi)} \, .
\label{eq:unitarity}
\end{eqnarray}

\section{Constraints and Benchmarks}
\label{sec:conbench}
From the discussion in the previous section and Eq.~(\ref{Eq:massdiff}), it follows that the mass-splitting parameter can take either positive or negative values depending on the value of $\lambda_{5}$.
Both these choice of $\lambda_{5}$ values which are compatible with the theoretical constraints that are illustrated in Eq.~(\ref{eq:stability}) and Eq.~(\ref{eq:unitarity}) respectively. Based upon the $\Delta M$ parameter one can expect three possible scenarios: (\emph{i}) \textbf{Degenerate Scenario} $(\Delta M \simeq 0)$: $m_{H^{\pm\pm}} \simeq m_{H^{\pm}} \simeq m_{H^{0}/A^{0}}$ (\emph{ii}) \textbf{Positive Hierarchy Scenario} $(\Delta M > 0)$: $m_{H^{\pm\pm}} > m_{H^{\pm}} > m_{H^{0}/A^{0}}$ and (\emph{iii}) \textbf{Negative Hierarchy Scenario} $(\Delta M < 0)$: $m_{H^{\pm\pm}} < m_{H^{\pm}} < m_{H^{0}/A^{0}}$. With these definitions in hand, we now elaborate on the phenomenological aspects of these BSM scalars. One can notice that all the potential parameters can be expressed in terms of the bare mass term\footnote{One can notice that the parameter $\lambda_{4}$ can be absorbed by redefining $\tilde{M}_{\Delta}$ while satisfying the identity $M^{2}_{\Delta}  = \tilde{M}^{2}_{\Delta} + \frac{\lambda_{4}}{2}v^{2}_{d}$.} ($M_{\Delta}$), $\Delta M$, $v_{t}$, and the mixing angle $\alpha$. For all practical purposes, we can consider these as the fundamental parameters to analyze the phenomenological features of the scalar sector for this BSM scenario. Here, we present the existing constraints on these parameters, which come from different experiments. 

\begin{itemize}
    \item[] To impose the bound on mixing angle $\alpha$, one can demand that the coupling between $h^{0}$ and other SM particles must remain within the SM expectation value. In addition, the presence of the additional charged scalars in this model can enhance the branching ratio of $h^{0} \to \gamma \gamma$ through loop-mediated contributions. However, to ensure that the properties of $h^{0}$ are compatible with the recent LHC measurements, the corresponding values of the mixing angle must satisfy the limit $\left|\sin\alpha\right| \lesssim 0.3$~\cite{Primulando:2019evb,Melfo:2011nx}.
    \item[] The electroweak precision observables, which are parametrised by the oblique coefficients ($S$, $T$, and $U$), receive additional contributions due to the presence of exotic scalar bosons. For this model, the current measurement on the $T$  parameter strictly disfavors the large mass difference between the triplet scalars. To satisfy this limit, one can impose the bound $\left|\Delta M\right| \lesssim 40$ GeV~\cite{Primulando:2019evb, Chun:2012jw,Aoki:2012jj,Dey:2020tfq}. For our numerical analysis, we take the best-fitted value for the S and T parameters provided
by the $\tt{Gfitter}$ Group: $S = 0.06 \pm 0.09$ and $T = 0.10 \pm 0.07$ with a correlation coefficient of +0.91~\cite{Baak:2014ora}.   
    \item[] The interaction Lagrangian as described in Eq.~(\ref{Eq:neutrinomass}) can give rise to different lepton flavor violating (LFV) processes, $\ell_{\alpha} \to \ell_{\beta}\gamma$ which are generated via both $H^{\pm}$ and $H^{\pm\pm}$ at 1-loop level. Along with that, the doubly charged scalar can participate in processes like $\ell_{\alpha} \to \ell_{\beta}\ell_{\gamma}\ell_{\delta}$ even at tree level. The upper limit on the branching fraction of these measurements enables us to determine the allowed region in the $v_{t} - M_{\Delta}$ parameter plane. We would like to point out that the mass splitting ($\Delta M$) between the triplet scalars does not directly affect the LFV processes and can be neglected. Following Refs.~\cite{Dev:2017ouk,Dinh:2012bp,Akeroyd:2009nu,Kakizaki:2003jk}, one can realise that the most stringent bound on the aforementioned parameter plane can be obtained while considering the $\mu \to e \gamma$~\cite{MEG:2016leq} and $\mu \to 3e$~\cite{SINDRUM:1987nra} processes. Satisfying the normal (inverted) mass hierarchy conditions, one can determine the following lower limit on $v_{t}$ depending upon the $m_{H^{\pm\pm}}$ mass -
    \begin{equation}
        v_{t} \gtrsim 0.78-1.5 \, (0.69 - 2.9)\times10^{-8}~\text{GeV}\times\frac{100~\text{GeV}}{m_{H^{\pm\pm}}} \,.
        \label{Eq:vtlimit}
    \end{equation}
\end{itemize}

We now present the updated collider limits for these triplet scalars, which correspond to three different mass hierarchy scenarios. As mentioned before, the triplet scalar masses can be parametrised by using two mass parameters $M_{\Delta}$ and $\Delta M$. We begin with negative and degenerate mass scenarios where the doubly charged scalars can decay either via pure leptonic mode ($H^{\pm\pm} \to \ell^{\pm}_{\alpha}\ell^{\pm}_{\beta}$) or via pure vector boson mode ($H^{\pm\pm} \to W^{\pm}W^{\pm}$)\footnote{For relatively lighter values of $m_{H^{\pm\pm}}$ one or both of these $W$ bosons can be off-shell.}. 

The vertices between the doubly charged scalar and the charged leptons are roughly proportional to $\sim\frac{m_{\nu}}{v_{t}}$. As a result, for $v_{t} \lesssim \mathcal{O}(10^{-4})$ GeV, these doubly charged scalars dominantly decay via leptonic mode with almost $100\%$ branching ratio. At the LHC, the $H^{\pm\pm}$ bosons can be pair-produced via the Drell–Yan process, leading to pairs of same-sign dilepton final states. This distinctive signature has motivated the experimental collaborations to perform several searches in this channel~\cite{ATLAS:2012hi,CMS:2012dun,ATLAS:2014kca,CMS:2016cpz,CMS:2017pet,ATLAS:2017xqs,CMS:2017fhs,ATLAS:2018ceg,ATLAS:2021jol,ATLAS:2022pbd}.
Non-observation of a significant excess \emph{w.r.t.} the SM expectations imposed a stringent bound on the $m_{H^{\pm\pm}}$. The ATLAS collaboration~\cite{ATLAS:2022pbd} has imposed a lower limit of $m_{H^{\pm\pm}} \lesssim 1020$ GeV for the $ H^{\pm\pm} \to \ell^{\pm}_{\alpha}\ell^{\pm}_{\alpha}$ final state where $\ell = e, \mu$. Similarly, the CMS collaboration \cite{CMS:2017pet} has determined the lower limit $m_{H^{\pm\pm}} \lesssim 535$ GeV for the $H^{\pm\pm} \to\tau^{\pm}\tau^{\pm}$ decay mode. 

On the other hand, for $v_{t} \gtrsim \mathcal{O}(10^{-4})$ GeV, the $H^{\pm\pm} \to W^{\pm}W^{\pm}$ decay mode opens up and in the range of $\mathcal{O}(10^{-3})~\text{GeV} \lesssim v_{t} \lesssim \mathcal{O}(1)$ GeV the corresponding branching ratio is almost $100\%$ for $\Delta M \leq 0$. In this scenario, the $H^{\pm\pm}$ scalars can decay into a pair of same-sign $W$ bosons, leading to characteristic collider signatures. Searches by the experimental collaborations~\cite{ATLAS:2018ceg,ATLAS:2021jol} have analysed these channels and excluded the mass range $200~\text{GeV} \lesssim m_{H^{\pm\pm}} \lesssim 350~\text{GeV}$, depending on the $W$ boson decay mode.

\begin{figure}
\includegraphics[scale=0.3]{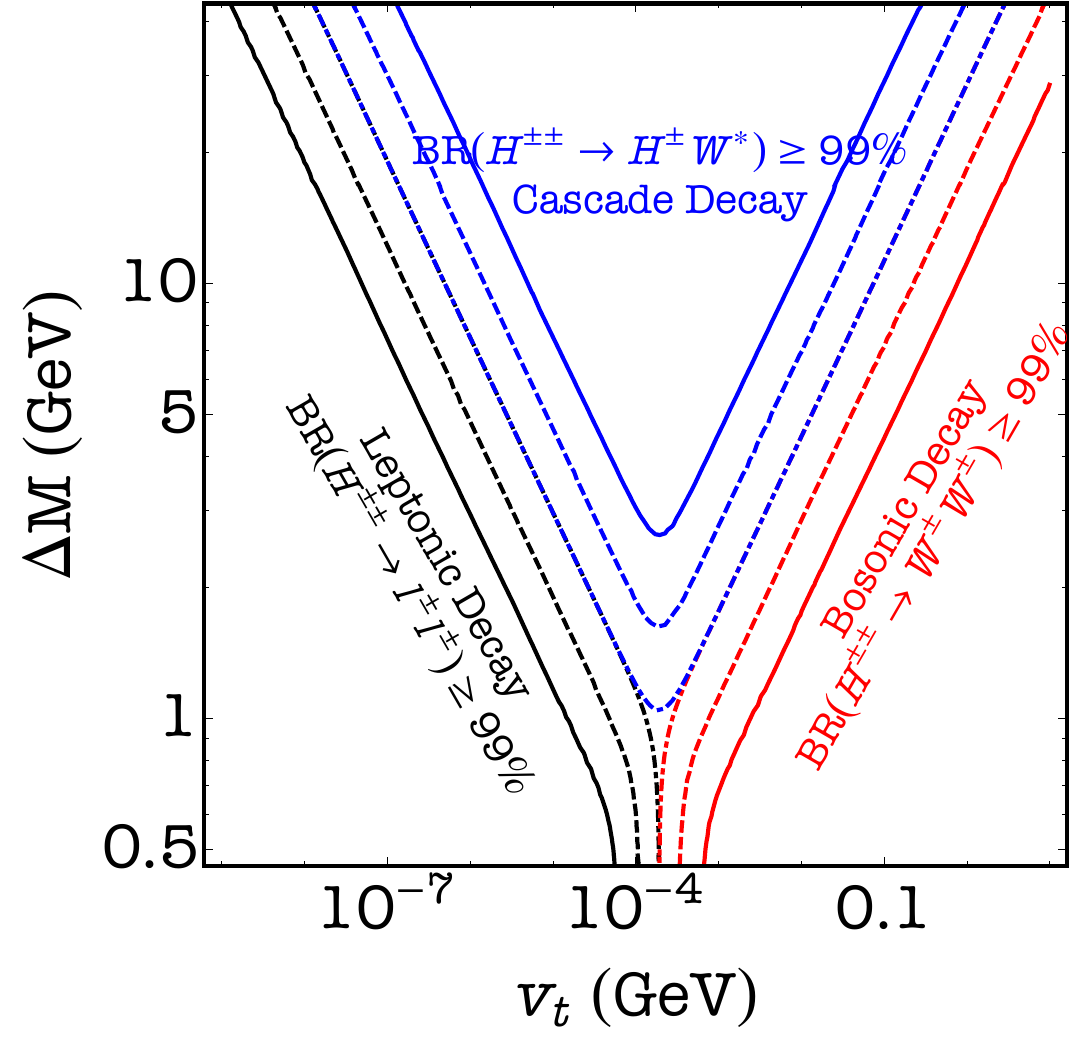}
\includegraphics[scale=0.3]{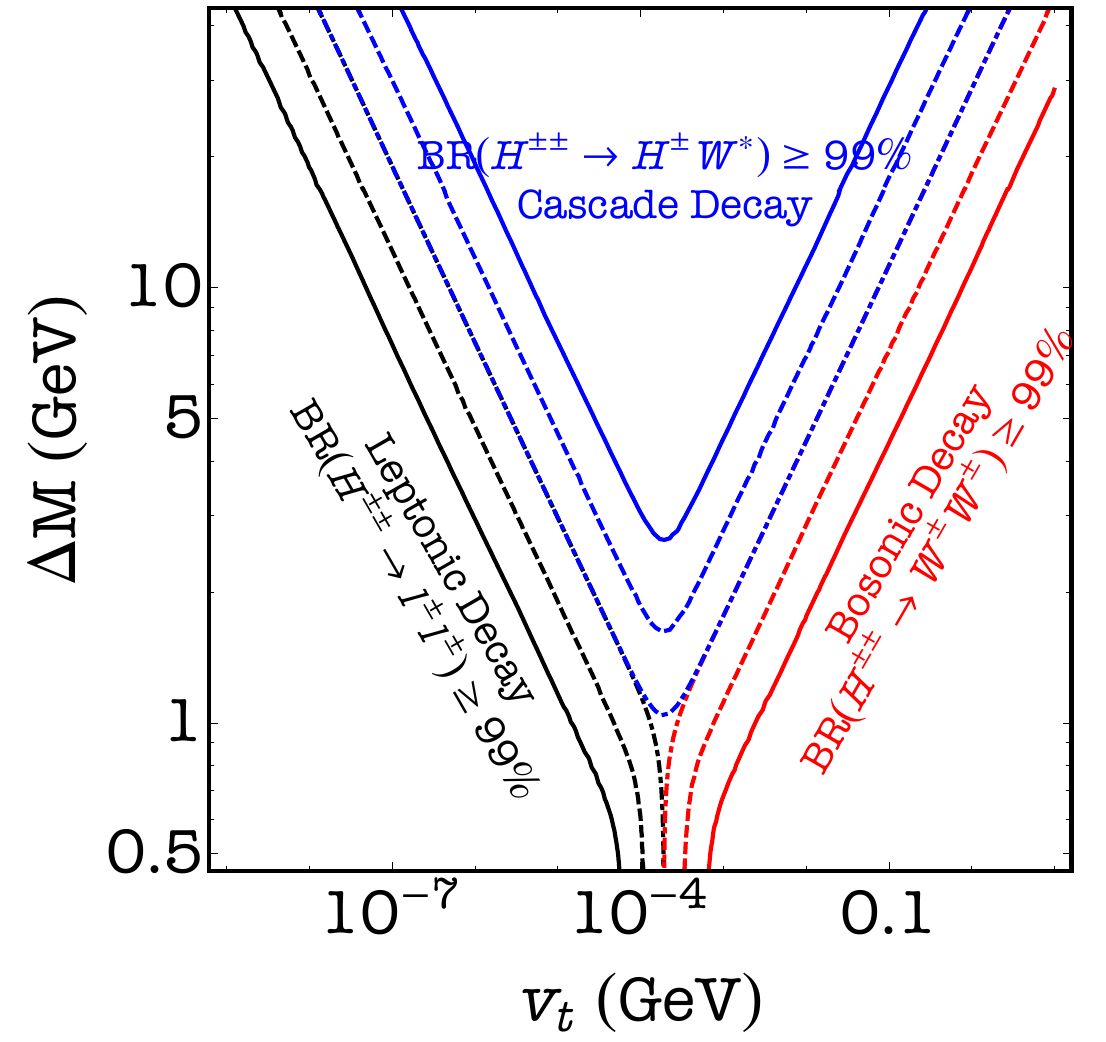}
\includegraphics[scale=0.3]{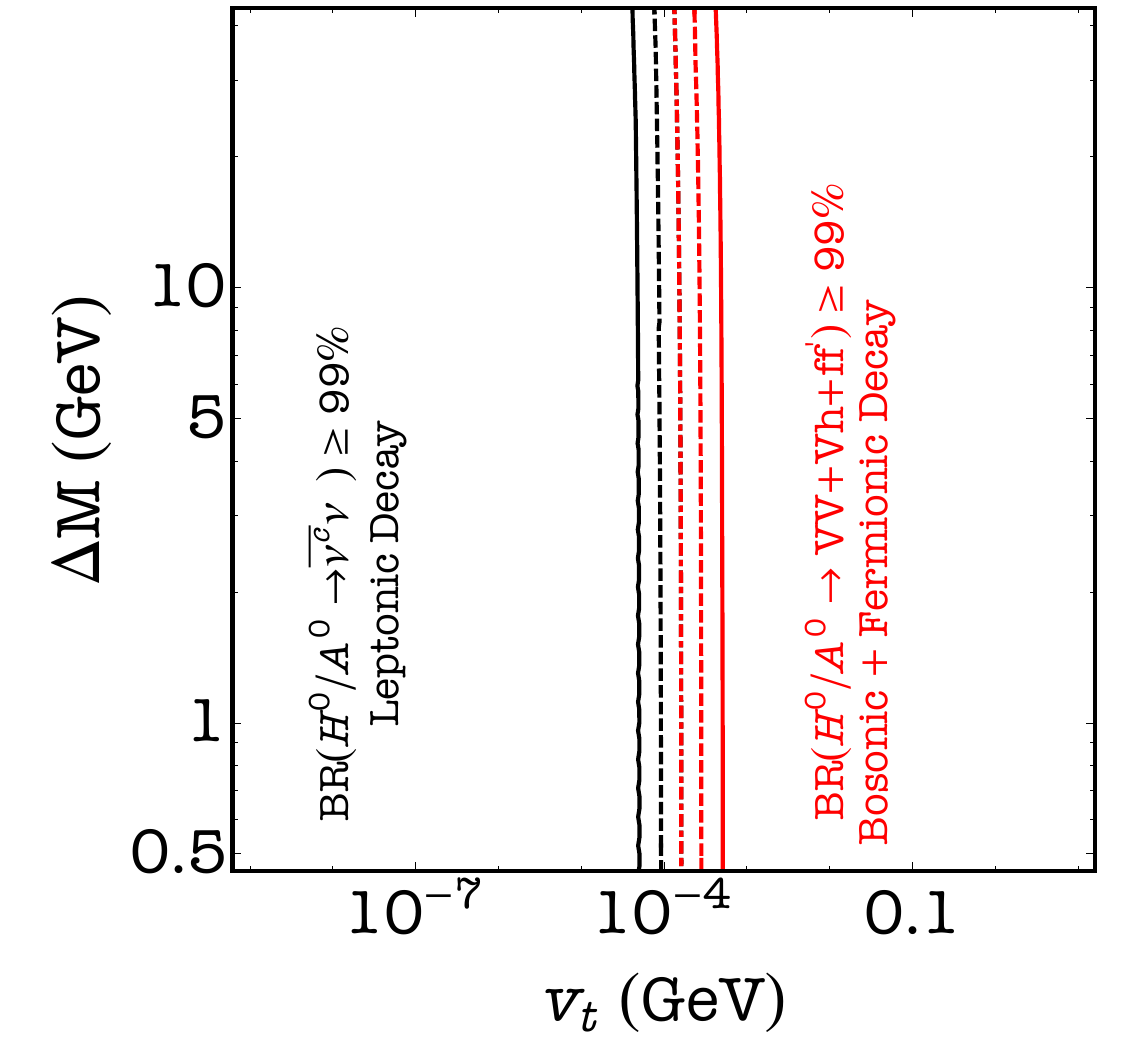}
\caption{Decay phase diagram of $H^{\pm\pm}$ (left), $H^{\pm}$ (middle), and $H^{0}/A^{0}$ (right) with $m_{H^{\pm\pm}}$ = 200 GeV for the positive mass hierarchy scenario. The solid, dashed and dot-dashed contours represent 99\%, 95\%, and 50\% branching ratios in different decay regions.}
\label{fig:BRexotic}
\end{figure}

However, the interpretation of the direct search limits significantly alters in the positive mass hierarchy scenario $\Delta M > 0$, where the cascade decay channels open up for both the doubly and singly charged scalars. To elaborate on this, we like to discuss the decays of triplet scalars in more detail. For $\Delta M > 0$, the doubly charged scalars can decay via three possible modes: \emph{(i)} leptonic decay, $H^{\pm\pm} \to \ell^{\pm}_{i}\ell^{\pm}_{j}$, \emph{(ii)} bosonic modes, $H^{\pm\pm} \to W^{\pm}W^{\pm}$ and \emph{(iii)} cascade mode, $H^{\pm\pm} \to H^{\pm}W^{*}$ and the ratio between their corresponding branching fractions can be represented approximately in the following manner:
\begin{equation}
    \ell\ell : WW : H^{\pm}W^{*} :: \left(\frac{m_{\nu}}{v_{t}}\right)^{2}\left(\frac{v_{d}}{m_{H^{\pm\pm}}}\right)^{2} : \left(\frac{v_{t}}{v_{d}}\right)^{2} : \frac{\text{max}(\Delta M, 0)^{5}}{v^{2}_{d}m^{3}_{H^{\pm\pm}}} \,.
    \label{Eq:hppbr}
\end{equation}

We would like to remind the reader that the mass gap between the triplet scalars has an upper limit of $\Delta M \lesssim 40$ GeV after satisfying the electroweak precision measurements. As a result, the on-shell production of the $W$ is kinematically forbidden. In Fig.~\ref{fig:BRexotic} \emph{left} panel, we present the branching ratio contours for the $H^{\pm\pm}$ in the $v_{t}$-$\Delta M$ plane, for the leptonic (black contours), bosonic (red contours), and cascade modes (blue contours), respectively. The solid line encloses the region where the corresponding BR is roughly around $\geq 99\%$. As expected for lower values of $v_{t}$, the doubly charged scalars can decay via leptonic mode, whereas for the large $v_{t}$ value, the bosonic decay modes dominate. Interestingly, for $\Delta M > 1$ GeV, the cascade decay mode opens up, and it becomes dominant in the region $\Delta M \gtrsim 10$ GeV and $10^{-5}~~\text{GeV} \lesssim v_{t} \lesssim 10^{-2}$ GeV. 

In the case of singly charged scalars, four possible decay modes open up in the positive mass hierarchy scenario \emph{i.e.} (\emph{i}) leptonic decay, $H^{\pm} \to \tau^{\pm} \nu$ \emph{(ii)} hadronic decay, $H^{\pm} \to t \bar{b}$, (\emph{iii}) diboson decay, $H^{\pm} \to W^{\pm}Z/h^{0}$ and \emph{(iv)} cascade decay, $H^{\pm} \to H^{0}/A^{0}W^{*}$ respectively. The ratio between the corresponding branching ratios for these channels can be expressed in the following manner:
\begin{align}
    \ell\nu : t\bar{b} : WZ :Wh^{0} : H^{0}/A^{0}W^{*} :: \left(\frac{m_{\nu}}{v_{t}}\right)^{2}\left(\frac{v_{d}}{m_{H^{\pm}}}\right)^{2} : \left(\frac{m_{t}}{v_{d}}\right)^{2}\left(\frac{v_{t}}{m_{H^{\pm}}}\right)^{2} : \left(\frac{v_{t}}{v_{d}}\right)^{2} : \left(1 - 2\zeta\right)^{2}\left(\frac{v_{t}}{v_{d}}\right)^{2}\nonumber \\
    : \frac{\text{max}(\Delta M, 0)^{5}}{v^{2}_{d}m^{3}_{H^{\pm}}} \, .
    \label{Eq:hcBR}
\end{align}
\noindent
Here $m_{t}$ is the top quark mass and $\zeta = \frac{v_{d}}{2v_{t}}\sin\alpha$. In Fig.~\ref{fig:BRexotic} \emph{central} panel, we illustrate the corresponding branching ratio contours for these channels. For the present discussion, we resort to a minimal mixing between the CP-even scalar states by setting the corresponding value of $\sin\alpha \approx \mathcal{O}(10^{-8})$. From Eq.~(\ref{Eq:hcBR}), one can realise that for a nearly mass-degenerate scenario in the region $v_{t} \lesssim 10^{-4}$ GeV, leptonic decay modes dominate (region enclosed by the solid black contour). Similarly, for the relatively large value of $v_{t}$ ($v_{t} \gtrsim 10^{-2}$ GeV), the combination of bosonic and hadronic decay modes surpasses other channels. However, for a moderate range of $v_{t}$ ($10^{-5}~~\text{GeV} \lesssim v_{t} \lesssim 10^{-3}$ GeV), the cascade decay modes become the primary channel for the mass gap larger than $\Delta M \gtrsim 10$ GeV. 

Apart from the charged scalars, both the $CP$-even and $CP$-odd scalars can decay into $\bar{\nu}^{c}\nu$ final state, which is described in the \emph{right} panel of Fig.~\ref{fig:BRexotic}. On the other hand, for the large $v_{t}$ value, these neutral scalars can either decay via bosonic or hadronic modes, depending upon their $CP$ properties as well as the coupling strength that governs those decay modes. 

Considering the decay of these triplet scalars, one can realise that for $\Delta M > 0$, there exists a region in the parameter plane that remains excluded from the existing direct search limits as the signal strength corresponding to both the leptonic and hadronic channels is suppressed. In fact, after conducting an elaborate analysis, Refs.~\cite{Ashanujjaman:2025scr,Ashanujjaman:2023tlj} have estimated that, in case of the Type-II seesaw model for $\Delta M \gtrsim 10$ GeV and triplet \emph{vev} ranging between $v_{t} \approx \mathcal{O}(10^{-7})~~\text{GeV} - \mathcal{O}(10^{-3})~~\text{GeV}$, remain available after considering all possible theoretical and experimental constraints. 

These particular findings serve as the central motivation for our study, where we propose a novel search strategy in the context of the LHC. In Table~\ref{table:benchmark}, we mention the corresponding values of relevant parameters that we would like to probe at LHC. As described, in this region of the parameter space, the charged triplet scalars primarily decay via cascade mode and generate off-shell gauge bosons and $H^{0}/A^{0}$. Due to the off-shell nature, the gauge bosons would further decay into leptons and jets with very low transverse momentum. Furthermore, in this unconstrained region, neutral triplet scalars would exclusively decay to $\bar{\nu}^{c}\nu$. As an outcome, the final state is populated with soft leptons, soft jets, and very low missing transverse momentum. This particular signature is difficult to probe as it fails to satisfy the triggering criteria of the existing detector. Our primary goal here is to formulate a novel search strategy that can possibly probe this challenging benchmark scenario. In the following section, we will demonstrate the details of our analysis strategy. 

Before concluding this section, we would like to remind the reader that for our analysis, $M_{\Delta}$, $\Delta M$, and $v_{t}$ are only parameters that are sufficient to describe the corresponding scalar phenomenology. For completeness, we set best-fit values for neutrino oscillation parameters that correspond to the normal hierarchy. As a result, the corresponding terms of the $Y_{\nu}$ matrix will be fixed accordingly. However, it is important to note that for our present analysis, these oscillation parameters do not play any noticeable role. 

\begin{table}[h!]
\centering
\begin{tabular}{|c|c|c|}
\hline
$M_{\Delta}$ [GeV]& $\Delta M$ [GeV] & $v_{t}$ [GeV] \\ [0.75ex]
     \hline 
     [170 GeV, 250 GeV] & [10 GeV, 30 GeV] & $10^{-4}$ GeV \\ [0.75ex]
     \hline
     \end{tabular}
     \caption{The range of scalar sector parameters that we consider to probe at the LHC.}
\label{table:benchmark}
\end{table}

\section{A cut-and-count analysis at the LHC}
\label{sec:cut&count}

In the context of the Type-II seesaw model, our primary objective in this study is to maximise the sensitivity in the $M_{\Delta}$–$\Delta M$ parameter plane, particularly in regions not yet constrained by existing searches. For that purpose, we therefore present a feasibility study for probing the scalar triplet states at the 14 TeV LHC.
As described before, we concentrate on a phenomenologically motivated region of parameter space where the doubly charged ($H^{\pm\pm}$) and singly charged scalars ($H^{\pm}$) decay predominantly via cascade modes - $H^{\pm\pm} \to H^{\pm} W^{*}$ and $H^{\pm} \to HW^{*}/AW^{*}$ respectively as described in Fig.~\ref{fig:BRexotic}. The exotic neutral scalars which are produced in these cascades subsequently decay through the invisible channel $H/A \to \bar{\nu}^{c}\nu$ mode. To ensure these decay patterns, we set the triplet scalar \emph{vev}, $v_{t} \simeq 10^{-4}$ GeV, and the mass gap between the triplet scalars ranges between 10-30 GeV. At the LHC, the triplet scalars can be inclusively produced via $p~p \to H^{\pm\pm}H^{\mp} \oplus H^{\pm\pm}H^{\mp\mp} \oplus H^{\pm}H^{\mp}$ modes. In the case of a compressed spectra scenario, the jets or the leptons which are produced from the off-shell $W$ decay are kinematically soft. Moreover, the neutral scalars produced in the $H^{\pm}$ decay predominantly yield invisible final states; the characteristic signal topology consists of missing transverse energy ($E^{miss}_{T}$) accompanied by soft leptons or jets. For a large fraction of signal events, these triplet scalars are pair-produced with negligible transverse boost. As a result, the missing transverse momentum, which is expected from these signal topologies, becomes negligible, which in turn makes this signal extraction from the corresponding SM background challenging. To overcome this difficulty, we chose the following signal process to develop our search strategy -
\begin{equation}
    p~p \to H^{\pm\pm}H^{\mp}j \oplus H^{\pm\pm}H^{\mp\mp}j \oplus H^{\pm}H^{\mp}j \, ,
    \label{Eq:signal}
\end{equation}
where the additional hard jet originates from ISR~\cite{Schwaller:2013baa,Baer:2014kya,Dutta:2017nqv}. The presence of this hard ISR jet imparts a substantial transverse recoil to the pair of charged triplet scalar system, thereby boosting decay products in the transverse plane. As a result, the otherwise soft and invisible decay products can collectively generate a sizeable missing transverse momentum, $E^{miss}_{T}$, which makes the signal experimentally accessible. As discussed, both the singly and doubly charged scalars will decay via cascade mode with a 100\% branching ratio and generate additional off-shell $ W$-bosons. For our analysis, we only consider the leptonic decay modes for these off-shell $W$ bosons. The primary reason to avoid the hadronic decay modes for these $W$s is that the jets arising from them fail to pass the reconstruction criteria for the high-luminosity LHC \footnote{The minimum transverse momentum threshold required for reliable jet reconstruction is $p_{T}(j) >$ 40 GeV.}. The dominant SM backgrounds that are relevant for our present analysis are 
\begin{itemize}
    \item $p~p \to t \bar{t} + \text{jets}$ \, ,
    \item $p~p \to W^{+}\ell^{-}\bar{\nu} + \text{jets} ~+ ~\text{h.c.}$ \, ,
    \item $p~p \to \tau^{+}\tau^{-} + \text{jets} $ \, ,
    \item $p~p \to W^{+}\ell^{+}\ell^{-} + \text{jets} ~+~ \text{h.c.}$ \, ,
    \item $p~p \to Z\ell^{+}\ell^{-} + \text{jets}$ \, ,
    \item  $p~p \to t W$ \, .
\end{itemize}
\noindent

In all these cases, the final-state neutrinos can give rise to the $E^{miss}_{T}$ spectrum, and $+jets$ denote both the one-jet and two-jet events. To generate the signal process, we incorporate the Type-II seesaw Lagrangian in $\tt{FEYNRULESv2.3.x}$~\cite{Alloul:2013bka} and develop the required $\tt{UFO}$ files. We use the $\tt{Madgraph5@NLOv2.9.21}$~\cite{Alwall:2014hca} to simulate both the signal and background events with the $\tt{NNPDF23~LO}$~\cite{Ball:2012cx} parton distribution function. At the parton-level event generation, we demand at least one jet with $p_{T}^{j} > 100$ GeV. This choice effectively suppresses low-energy QCD backgrounds while simultaneously providing a substantial transverse recoil to the charged triplet scalar system. All the cross sections are calculated at the tree level. After the parton-level generation, the simulated events are passed to $\tt{PYTHIA8}$~\cite{Bierlich:2022pfr} to incorporate the showering and hadronization effects. The detector level simulation is performed using $\tt{DELPHESv3.5.0}$~\cite{deFavereau:2013fsa} with the help of an appropriate card that enable us to use low $p_{T}$ leptons. To evade the double counting, we use the MLM scheme for the jet matching and merging~\cite{Mangano:2006rw}. To reconstruct various objects that are present in the final state, we use the following criteria. 
\begin{itemize}
    \item[] Jets are reconstructed using the $\tt{Fastjet}$ package~\cite{Cacciari:2011ma} with the anti-$k_T$ algorithm and jet radius parameter $R = 0.4$. All jets are required to be hadronic clusters satisfying $|\eta| < 5$ and $p_T > 40$ GeV.
    \item[] In the present analysis, the final signal state consists of kinematically soft leptons ($\ell = e, \mu$) satisfying $|\eta| < 2.5$ and $p_{\ell} \geq 5$ GeV. The tagging efficiencies are taken from the detector card $\tt{CMS\_PhaseII\_0PU.tcl}$ provided with $\tt{DELPHESv3.5.0}$.
    \item[] Hadronic clusters are identified as $b$-jets if they contain a $B$-hadron with $p_T(b) > 20$ GeV and $|\eta| \leq 3.4$. The $b$-tagging and misidentification efficiencies are taken from the card $\texttt{btagMedium.tcl}$.
    \item[] Similarly, hadronic tau leptons ($\tau_h$) are identified with the kinematic requirements $p_T(\tau_h) > 20$ GeV and $|\eta(\tau_h)| < 2.3$. The tau-tagging efficiencies are implemented using the parameterized functions of $p_T$ and $\eta$ provided in the card $\tt{CMS\_PhaseII\_0PU.tcl}$.
    \end{itemize}
    
    

Finally, to calculate the statistical significance for our analysis, we use the following formula 
\begin{equation}
    \mathcal{Z} = \frac{\sqrt{\mathcal{L}}\sigma_{S}}{\sqrt{\sigma_{S}+\sigma_{B}}} \, .
    \label{eq:signi}
\end{equation}

We begin our analysis by employing a $b$-quark veto cut where we assume medium tagging efficiency consistent with current LHC detector performance. This veto plays an important role in rejecting those backgrounds which have top quarks in the final state. With this cut, we can reject approximately one-third of top-associated background events while nominally affecting the signal events, which typically fail to contain heavy flavor quarks. In a similar spirit, we impose a veto on hadronically decaying $\tau_{h}$. Using this veto, we can suppress around 40\% of the tau-pair background events while marginally affecting signal and other background processes. Furthermore, we select those events that consist of a single jet in the final state topology. These particular selection cuts significantly reduce the $t\bar{t}$+jets and $W\ell\nu$+jets background as highlighted in Table~\ref{tab:deltaM10}. With these selection criteria in hand, we now consider different kinematic distributions to further improve our signal background efficiency. 
\begin{figure}
    \includegraphics[scale=0.3]{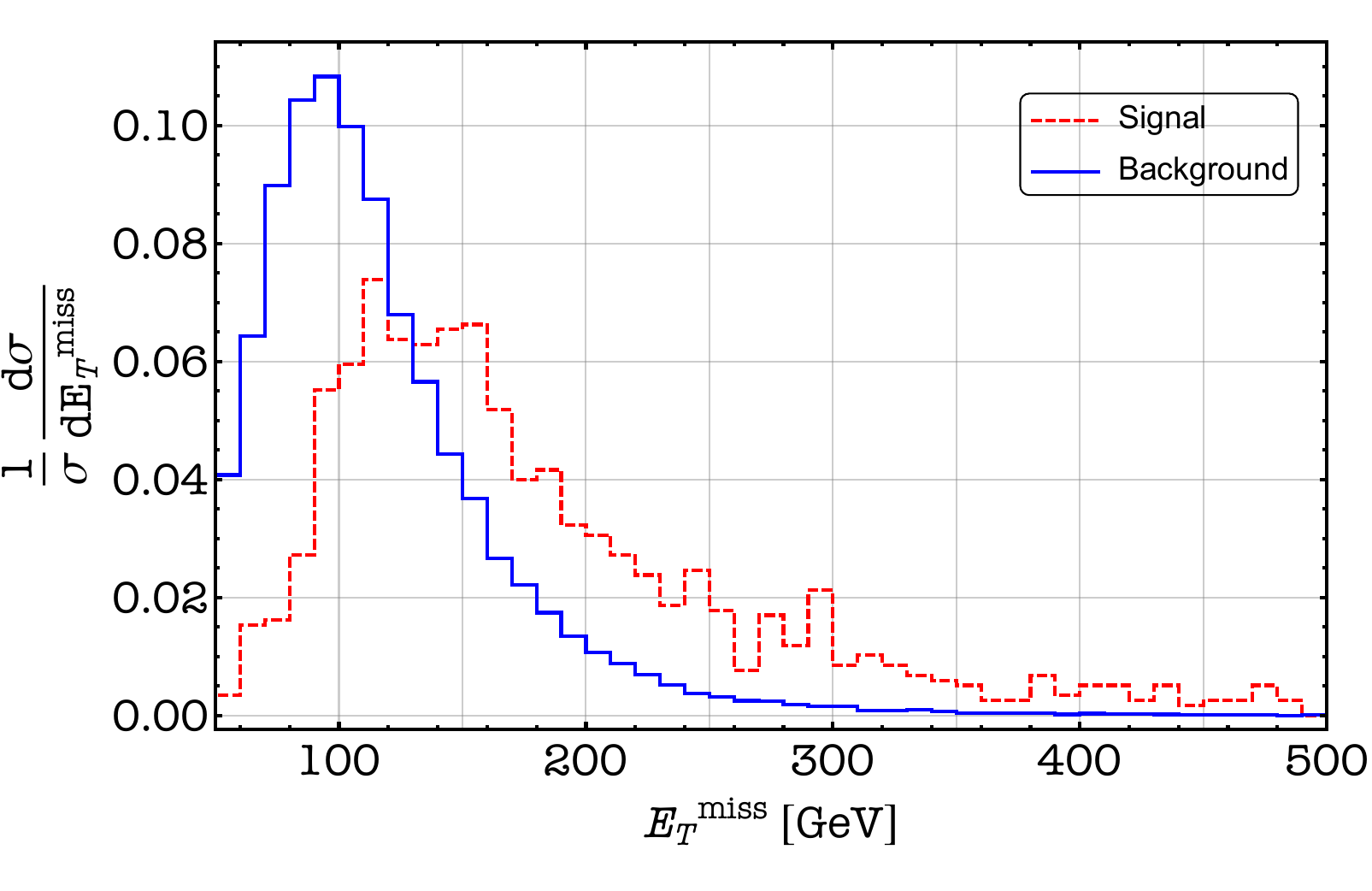}
    \includegraphics[scale=0.3]{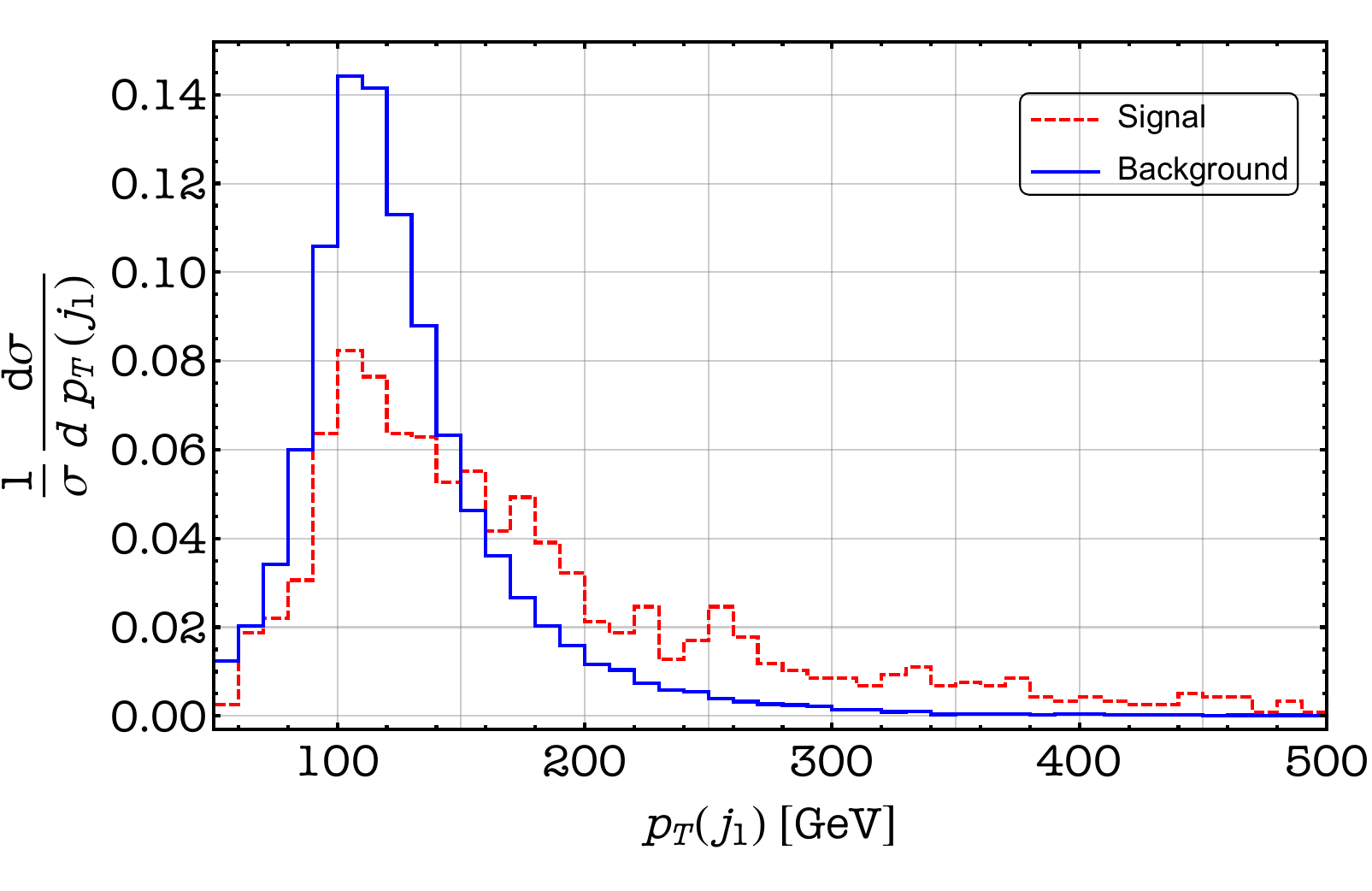}
\caption{Normalized differential distribution corresponds to missing transverse momentum $E^{miss}_{T}$ (left) and the leading transverse momentum of the jet $p_{T}(j_{1})$ for the signal and the total SM background. The solid blue line and the red dashed line represent the signal and the total SM background, respectively.} 
\label{fig:pTj1ETmiss}
\end{figure}

In Fig.~\ref{fig:pTj1ETmiss}, we demonstrate the normalized  distribution correspond to $E^{miss}_{T}$ (\emph{left}) and leading jet $p_{T}(j_{1})$ (\emph{right}) respectively. At the generation level, we have imposed an inclusive criterion with $p_{T}^{j} > 100$ GeV. As a result, one can notice that the majority of signal and background events are distributed in the high $p_{T}$ value. In contrast, if we consider the $E^{miss}_{T}$ distribution, for the signal events the emitted hard jet would give rise to the large missing transverse energy for the triplet-pair system. As a result, one can find a similarity between the shape of $E^{miss}_{T}$ and $p_{T}(j_{1})$ for the signal. On the other hand, this correlation between both these distributions does not appear for the SM backgrounds, which motivates us to devise the following cuts to distinguish the signal from backgrounds 
\begin{equation}
    E^{miss}_{T} > 120~ \text{GeV} ~ \& ~ p_{T}\left(j_{1}\right) > 120~ \text{GeV with}~ \left|\eta\left(j_{1}\right)\right| \leq 2.5.
\end{equation}

After applying the aforementioned kinematic cuts, we further demand the presence of two leptons in the final state. We like to point out that the lepton tagging efficiency primarily depends on their corresponding transverse momentum, and if it's $p_{T}^{\ell} < 5$ GeV, then corresponding tagging efficiency is zero. In addition to the range 5 GeV $< p_{T}^{\ell} <$ 10 GeV, the tagging efficiency roughly varies between 50\% to 70\% depending on detector performance. As a result by demanding these di-lepton ($N_{\ell}$ = 2) systems in practice, only 10\% of the signal events survive. Nevertheless, this requirement is essential to suppress purely hadronic backgrounds and to maintain a clean experimental signature. 

\begin{figure}
\centering
    \includegraphics[scale=0.3]{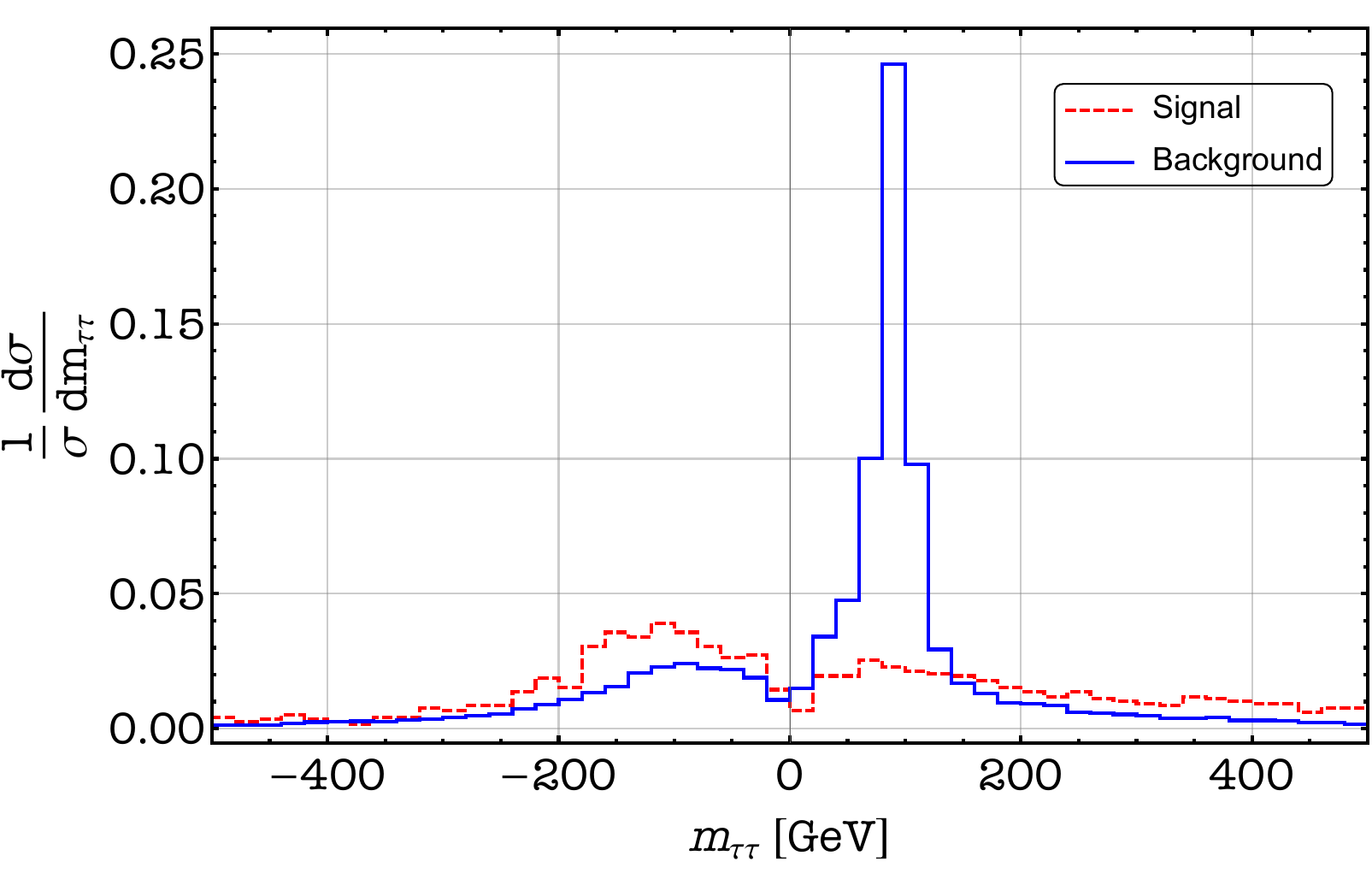}
\caption{Normalized differential distribution corresponds to $m_{\tau\tau}$ variables for the signal and the total SM background. The solid blue line and the red dashed line represent the signal and the total SM background, respectively.}
\label{fig:mtata}
\end{figure}

At this stage of analysis, the $\tau^{+}\tau^{-}$+jets process emerges as the dominant SM background, which makes the signal extraction from the background challenging. Hence, we adopt the di-tau invariant mass observable, which was described in Ref.~\cite{Dutta:2017nqv}. Before defining this variable, we'd like to elaborate on the $\tau^{+}\tau^{-}$+jets background. For these SM processes, both these $\tau$ pairs are produced due to the decay of highly boosted $Z$-bosons. On top of that, we assume that both these tau lepton is further decay via pure leptonic mode $\tau \to \ell \nu \bar{\nu}$. Since the parent $\tau$'s are produced with an ultra-relativistic momentum, all the daughter particles that arise from the $\tau$ decays would be expected to be collimated towards the parent particle's direction. Additionally, one can expect that the neutrinos produced from the individual $\tau$ decay carry the momentum of $\xi_{1}p_{T}(\ell_{1})$ and $\xi_{2}p_{T}(\ell_{2})$ respectively. Here, both the $\xi_{i}$ represent the momentum fraction carried by the neutrinos towards the transverse plane and $\ell_{i}$ are the leptons produced from each of the parent $\tau$. Considering the momentum conservation in the transverse plane for the $Zj$ system, we can write down the following identity: 
\begin{equation}
    -p_{T}(j_{1}) = (1 + \xi_{1})p_{T}(\ell_{1}) + (1 + \xi_{2})p_{T}(\ell_{2}) \, .
    \label{Eq:momen}
\end{equation}
In principle, Eq.~(\ref{Eq:momen}) comprises independent equations, and using the measured value of $p_{T}(j_{1})$, it is possible to calculate the actual value $\xi_{i}$ for the event-by-event level. With the collinear tau decay approximation, the corresponding invariant mass for the di-tau system can be expressed as 
\begin{equation}
    m^{2}_{\tau\tau} = m^{2}_{\ell_{1}\ell_{2}}(1 + \xi_{1})(1 + \xi_{2}) \,
    \label{Eq:mtata}
\end{equation}
where $m^{2}_{\ell_{1}\ell_{2}}$ signifies the invariant squared mass of leading and sub-leading leptons. In Fig.~\ref{fig:mtata}, we illustrate the normalized distribution of $m_{\tau\tau}$ variable which is defined as $m_{\tau\tau} \equiv \text{Sign}[m^{2}_{\tau\tau}]\times\sqrt{|m^{2}_{\tau\tau}|}$. For the background events, one can notice a sharp peak at the $Z$-boson pole mass value, which arises from the di-tau system. In contrast, for the signal and other SM backgrounds, the $E^{miss}_{T}$ direction and leptons' momentum direction are uncorrelated, and the events are distributed over the entire range of $m_{\tau\tau}$. Our primary objective here is to reject all those events which are from the $Z$ boson decay. Along with that, to achieve further sensitivity, we only select those events that satisfy the following criteria: 
\begin{equation}
       m_{\tau\tau} > 200~\text{GeV} ~\text{and} ~ m_{\tau\tau} < -100~\text{GeV}
    \label{Eq:cutmtata}
\end{equation}
With this cut, one can reject the majority of the $\tau\tau$+jets events, which dominantly reside around the $Z$-boson mass pole. In addition, the $t\bar{t}$+jets events are also lying around $-100~\text{GeV}~< m_{\tau\tau} < 200$ GeV bins. These guide us to devise an asymmetrical cut for this variable. 

\begin{figure*}
    \includegraphics[scale=0.3]{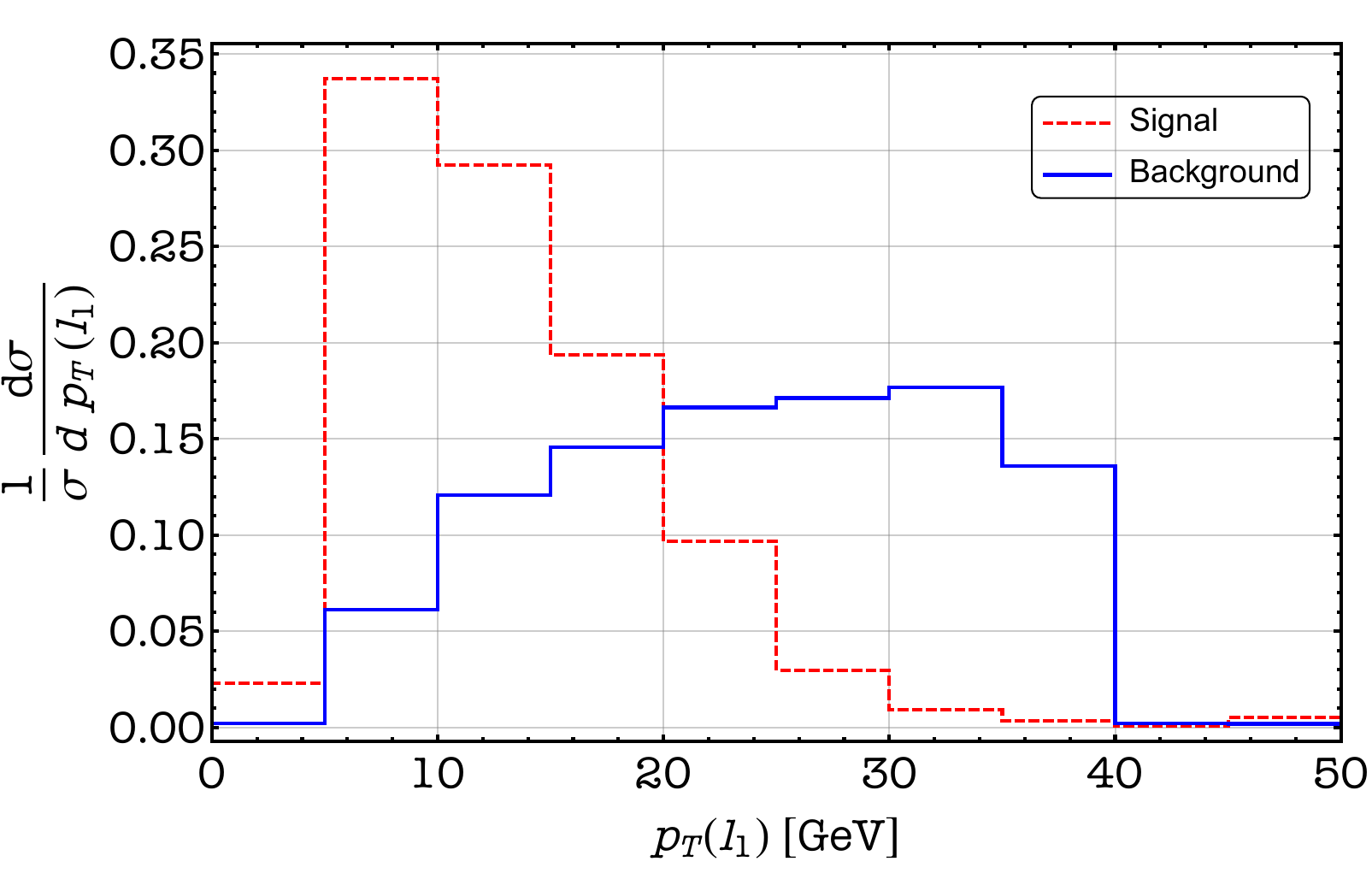}
    \includegraphics[scale=0.3]{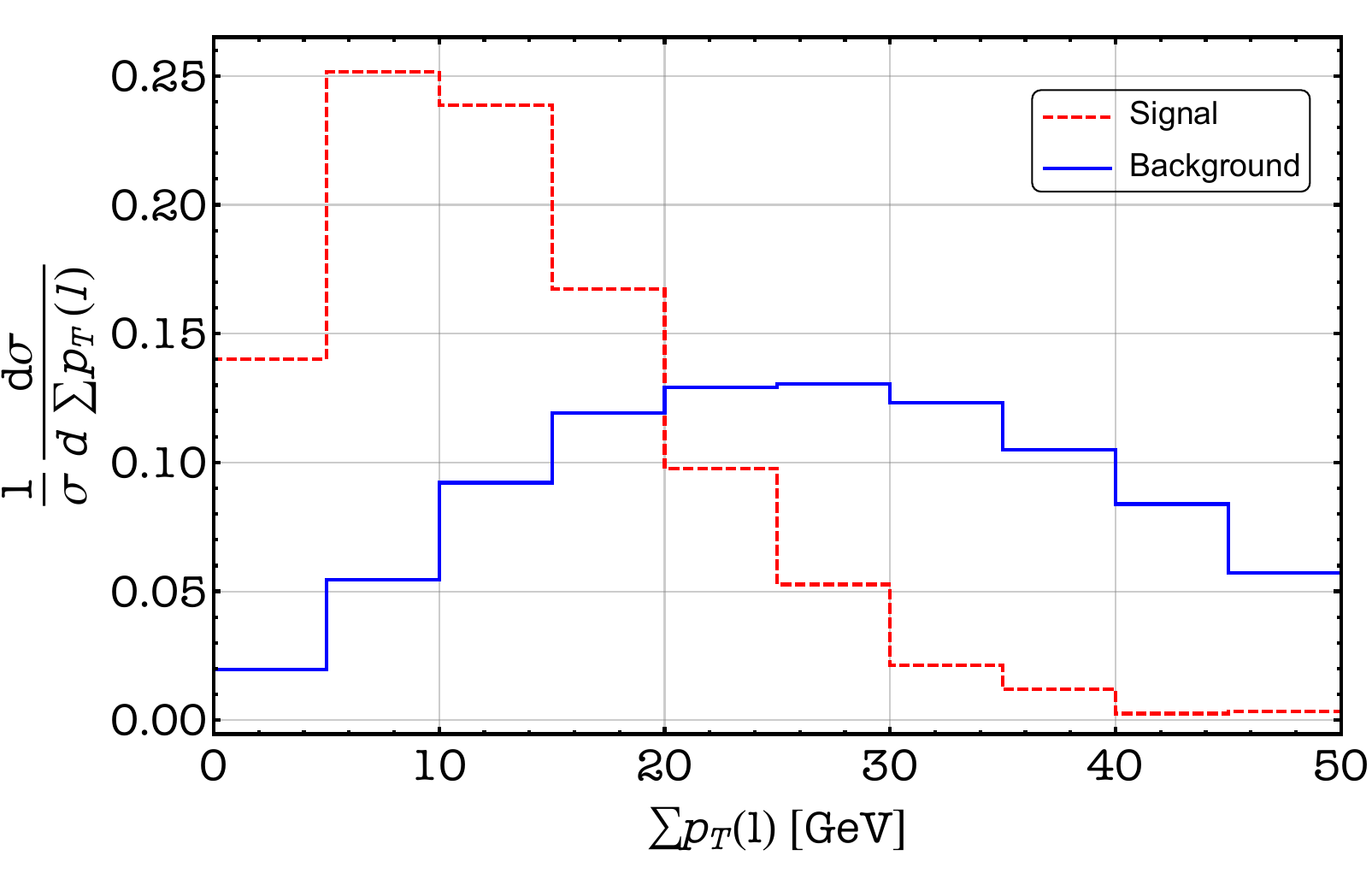}
\caption{Normalized differential distribution corresponds to the leading transverse momentum of the jet $p_{T}(\ell_{1})$ (left) and the vector sum of lepton transverse momentum $\sum_{i= 1, 2}p_{T}(\ell_{i})$  for the signal and the total SM background. The solid blue line represents the total SM background, and the red dashed line represents the signal, respectively.}
\label{fig:pTlsumpt}
\end{figure*}

After implementing all these cuts, the total number of surviving signal events is substantially reduced. At this point, our objective is to find suitable kinematic variables that can suppress the residual background without significantly affecting the signal events. To achieve this goal, we like to examine the kinematics of the leptons that are present in the final state. In the case of the signal, the leptons are produced due to subsequent decays of off-shell $W$-bosons. For moderately small values of $\Delta M$, leptons are expected to be relatively soft. In Fig.~\ref{fig:pTlsumpt}, we present the normalized distribution corresponding to the leading lepton $p_{T}(\ell_{1})$ (\emph{left}) and vector sum of leptons, $\sum p_{T}(\ell)$, respectively. As expected, for both these variables, the majority of the signal events populate the low $p_{T}$ region. In contrast, for the dominant backgrounds, the leptons are typically produced from on-shell gauge bosons and therefore exhibit harder transverse momentum spectra. Motivated by the distributions shown in Fig.~\ref{fig:pTlsumpt}, we therefore impose the following upper bounds on the transverse momentum of the leading lepton, $p_{T}(\ell)$, as well as on the scalar sum of lepton transverse momenta, $\sum p_{T}(\ell)$ 
\begin{equation}
    p_{T}\left(\ell_{1}\right) <  20 ~\text{GeV} ~~~~~~~~ \text{and} ~~~~~~~~~ \sum_{i =1,2}p_{T}\left(\ell_{i}\right) <  25 ~\text{GeV}  . 
\end{equation}
These requirements significantly suppress the residual backgrounds while retaining a sizable fraction of the signal events, particularly in the compressed mass spectrum regime under consideration. We emphasize that these cuts are specifically designed to target the soft-lepton signature intrinsic to the cascade decay topology of the Type-II seesaw scalars and are complementary to the previously imposed $E^{miss}_{T}$, ISR jet, and $m_{\tau\tau}$ selection.

\begin{table}[htb!]
    \begin{center}
            \begin{tabular}{|m{5em}||c|}
    \hline
     Cut-0    & Initial cross section   \\ [0.5ex]
     \hline
     Cut-1    & $b$-quark veto \\ [0.5ex]
     \hline
     Cut-2    & $\tau_{h}$ veto \\ [0.5ex]
     \hline
     Cut-3    & $N_{j}$ = 1 \\ [0.5ex]
     \hline
     Cut-4    & $p_{T}\left(j_{1}\right) > 120$GeV  and $\left|\eta\left(j_{1}\right)\right|$ \\ [0.5ex]
     \hline 
     Cut-5 &   $E^{miss}_{T} > 120$ GeV \\ [0.5ex]
     \hline
     Cut-6 & $N_{\ell}$ = 2 \\ [0.5ex]
     \hline
     Cut-7 & $m_{\tau\tau} >$ 200 GeV and $m_{\tau\tau} <$ -100 GeV \\ [0.5ex]
     \hline
     Cut-8 & $p_{T}\left(\ell_{1}\right) < $ 20 GeV \\ [0.5ex]
     \hline
     Cut-9 & $\sum_{i =1,2}p_{T}\left(\ell_{i}\right) < $ 25 GeV \\ [0.5ex]
     \hline
    \end{tabular}
        \end{center}
    \caption{Definition of cuts that are implemented in the signal background analysis.}
    \label{tab:placeholder}
\end{table}

The cumulative effect of all selection cuts is summarized in Table~\ref{tab:deltaM10} for $\Delta M =$ 10 GeV and 30 GeV, respectively, where we present the cut-flow for both the signal and the relevant SM backgrounds. As evident from this table, the combination of all the cuts leads to a substantial suppression of the background while maintaining a non-negligible signal yield. In particular, the $\tau^{+}\tau^{-}$+jets and $t\bar{t}$+jets backgrounds are efficiently reduced at successive stages of the analysis, resulting in a significantly improved signal-to-background ratio in the final selection.
\begin{table}[htb!]
    \centering
    \setlength{\tabcolsep}{0.5pt}
    \begin{tabular}{|c||c|c|c|c|c|c|c|c|}
    \hline 
         & (200, 10) [GeV] & (200, 30) [GeV]  & $t\bar{t}+\text{jets}$ & $\tau\tau+\text{jets}$ & $WW+\text{jets}$ & $Z\ell\ell+\text{jets}$ & $W\ell\ell+\text{jets}$ & tW \\ 
          & [fb] & [fb] & [fb] & [fb] & [fb] & [fb] & [fb] & [fb]\\
         \hline 
         Cut-0 & 25.55 & 20 & $4.0214\times10^{4}$ & $2.2237\times10^{4}$ & $3.267\times10^{3}$ & 107.07 & 390.219 & $1.42\times10^{4}$ \\
           \hline 
         Cut-1 & 23.61 & 17.746 & $9.57\times10^{3}$ & $2.0772\times10^{4}$ & $2.945\times10^{3}$ & 87.56 & 347.49 & $5.95\times10^{3}$ \\
        \hline
        Cut-2 & 23.038 & 16.576 & $8.847\times10^{3}$ & $1.2576\times10^{4}$ & $2.782\times10^{3}$ & 82.486 & 324.814 & $5.67\times10^{3}$ \\
        \hline 
       Cut-3 & 9.543 & 3.79 & 99.433 & $2.111\times10^{3}$ & 113.92 & 0.9054 & 2.299 & $1.08\times10^{3}$ \\
         \hline   
      Cut-4  & 5.668 & 2.13 & 45.4503 & 646.958 & 50.248 & 0.405 & 0.9402 & 50.065\\
         \hline
        Cut-5 & 5.282 & 1.862 & 32.8601 & 284.715 & 39.078 & 0.3417 & 0.5916 & 30.97 \\
        \hline
        Cut-6 & 0.363 & 0.364 & 12.3326 & 47.044 & 12.952 & 0.2394 & 0.2766 & 10.8775 \\
        \hline
 Cut-7 & 0.2606 & 0.272 & 9.9981 & 4.567 & 10.168 & 0.141 & 0.1878 & 4.9875 \\
     \hline
    Cut-8  & 0.2135 & 0.1396 & 2.7209 & 2.0416 & 2.2617 & 0.0561 & 0.0792 & 0.9975 \\
       \hline
    Cut-9  & 0.2044 & 0.1304 & 2.3023 & 1.8304 & 1.827 & 0.048 & 0.0714 & 0.855\\
       \hline
    \end{tabular}
    \caption{Summary of effective cross-section for the signal and dominant background at a 14 TeV $pp$ collider. For the signal process, we set the singly charged scalar mass at $M_{\Delta}$ = 200 GeV and the triplet scalar mass gap $\Delta M$ = 10 GeV and 30 GeV. The second row highlights the initial cross section after demanding at least one jet with $p_{T}^{j} \geq$ 100 GeV and MLM matching.}
    \label{tab:deltaM10}
\end{table}

\begin{figure}[htb!]
\centering
    \includegraphics[scale=0.5]{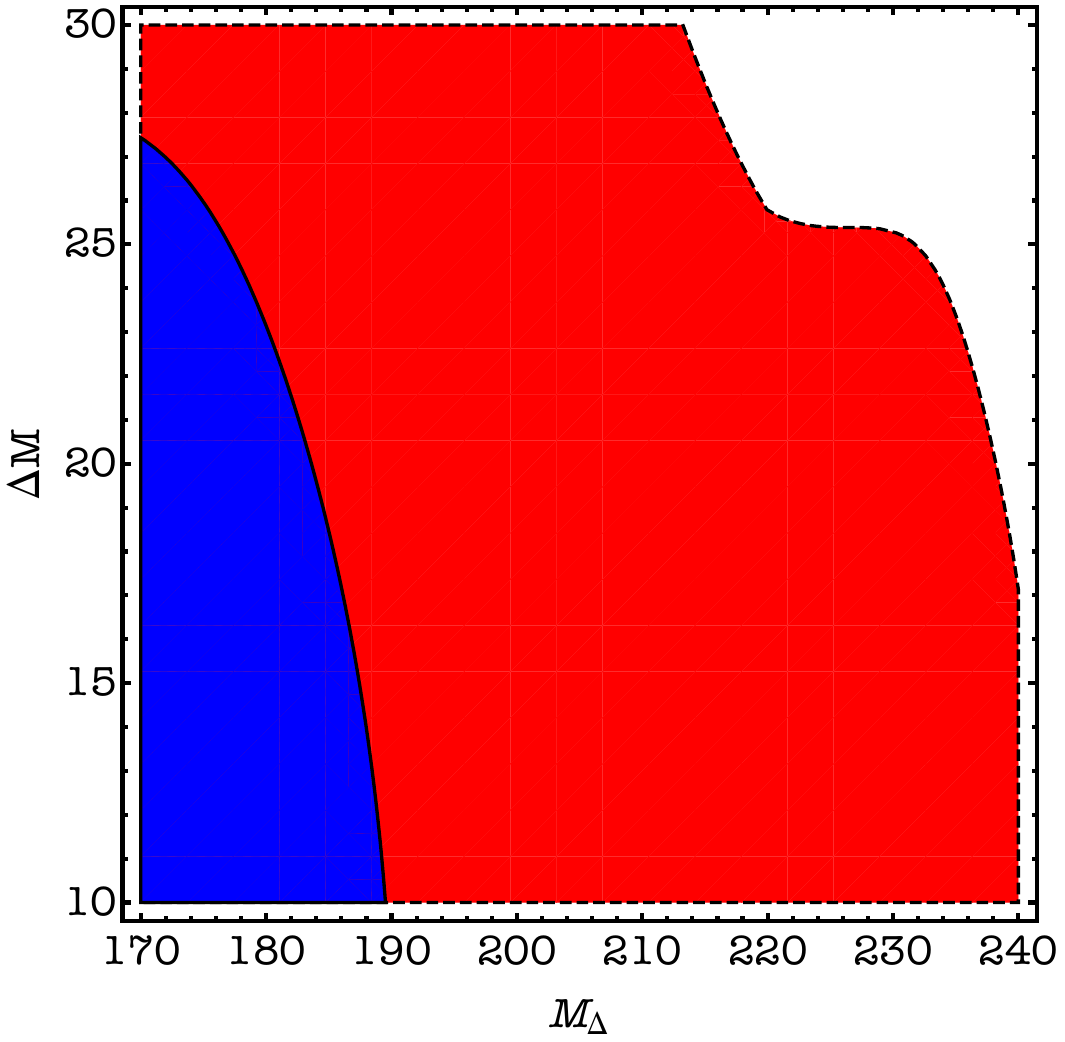}
\caption{Projected sensitivity in the $M_{\Delta} - \Delta M$ plane after applying the cut-based analysis of the signal and background distribution at the 14 TeV LHC. The black dashed and solid black curves correspond to 2$\sigma$ exclusion and 5$\sigma$ discovery contours, respectively, assuming an integrated luminosity of $\mathcal{L} = 3000~fb^{-1}$. The shaded region indicates the parameter space where signal significance exceeds the corresponding confidence level, with the red (blue) shaded region representing 2$\sigma$ (5$\sigma$) sensitivity reaches. The unshaded region can not be probed using the present analysis assumptions.}
\label{fig:exclusion}
\end{figure}

Using the final event yields after all cuts, we evaluate the statistical significance for different combinations of $\Delta M$ and $M_{\Delta}$ values following the prescription described in Eq.~(\ref{eq:signi}). Based on this analysis, we derive the projected sensitivity at the 14 TeV LHC in the $M_{\Delta} - \Delta M$ parameter plane as shown in Fig.~\ref{fig:exclusion} for integrated luminosity $\mathcal{L} = 3000~fb^{-1}$. The blue shaded region represents $5\sigma$ discovery, and the red shaded region represents $2\sigma$ exclusion reach for the proposed analysis.  As evident from this figure, our strategy exhibits significant sensitivity in the moderately compressed regime, particularly for mass splitting in the range 10 GeV $\lesssim \Delta M \lesssim$ 30 GeV. The reach extends up to $M_{\Delta} \sim $ 220-230 GeV at the 2$\sigma$ level, while a robust $5\sigma$ discovery is achievable for comparatively lower triplet masses roughly around $M_{\Delta} \lesssim$ 190 GeV, depending on the exact value of $\Delta M$. We like to remind you that for this analysis we set the triplet \emph{vev} at $v_{t} = 10^{-4}$ GeV. The sensitivity degrades for very small mass splitting, where the decay products become extremely soft and fail to satisfy the lepton reconstruction thresholds, as well as for larger $\Delta M$, where the leptons become harder, and the signal begins to resemble SM backgrounds more closely. In addition, for a large value of $M_{\Delta}$, the initial signal cross section, it is significantly low to achieve desirable sensitivity. 
This demonstrates that ISR-assisted searches provide a powerful and complementary avenue to explore compressed triplet scalar spectra in the Type-II seesaw model. The results highlight the importance of dedicated analyses targeting such challenging regions of parameter space, which may otherwise remain elusive in standard high-$p_{T}$
searches.

\section{Summary and Conclusions}
\label{sec:concl}


In this work, we have revisited the collider phenomenology of the Type-II seesaw model with a particular focus on regions of parameter space that remain weakly constrained by existing experimental searches. Current limits on the mass of the doubly charged Higgs boson are largely derived under the assumption that its decays are dominated either by same-sign dilepton or same-sign diboson final states. However, as emphasized in recent studies and demonstrated in this analysis, such assumptions are not generic once moderate mass splittings among the triplet scalars are taken into account. In the presence of compressed spectra, cascade decays such as these $H^{\pm\pm} \to H^{\pm}W^{\pm}$ become dominant, while the conventional leptonic and bosonic decay modes are suppressed.

We have shown that this suppression arises from two intertwined effects: the reduced phase space available for direct decays of the doubly charged scalar and the softness of the visible decay products emerging from the cascade chains. In particular, for moderate mass splitting of 10 GeV $\lesssim \Delta M \lesssim$ 30 GeV and triplet vev ranging between $10^{-5}~\text{GeV} \lesssim v_{t} \lesssim 10^{-3}$ \text{GeV}, the decay widths of $H^{\pm\pm} \to \ell^{\pm} \ell^{\pm}$ and $H^{\pm\pm} \to W^{\pm} W^{\pm}$ become subdominant. At the same time, the subsequent decays of the charged and neutral triplet scalars predominantly yield soft leptons and missing transverse energy, rendering standard search strategies largely ineffective. Consequently, sizeable regions of the Type-II seesaw parameter space remain unconstrained despite stringent bounds reported by the ATLAS and CMS collaborations.

Motivated by these challenges, we proposed a dedicated collider search strategy specifically tailored to this elusive regime. Our analysis exploits the presence of a hard initial-state radiation jet to boost the triplet scalar system, thereby generating a detectable missing transverse momentum signature. The signal is characterized by missing energy accompanied by soft leptons originating from off-shell $W$ bosons produced in the cascade decays of the charged scalars, as well as neutrino-dominated decays of the neutral states. By combining object-level vetoes, kinematic selections, an asymmetric di-tau invariant mass cut to suppress the $\tau^{+}\tau^{-} +$ jets background, and requirements exploiting the softness of the leptons, we demonstrated that the signal can be efficiently separated from the dominant SM backgrounds within a cut-and-count framework.

Our results show that, at the 14 TeV LHC, meaningful evidence-level and discovery-level sensitivities can be achieved in regions of the $M_{\Delta}- \Delta M$ parameter plane that are currently poorly constrained. In particular, we can achieve a 5$\sigma$ discovery sensitivity for triplet scalar masses in the range 170 GeV - 190 GeV, with mass gap varying between 10 GeV and 25 GeV. This proposed strategy significantly extends the experimental reach into the moderately compressed regime, highlighting its complementarity to existing searches optimized for hard leptonic or bosonic final states. Overall, this study underscores the importance of accounting for cascade-dominated decay topologies and compressed spectra when interpreting collider limits on extended scalar sectors. Dedicated searches of the kind presented here will be essential for fully probing the Type-II seesaw model and, more broadly, for uncovering collider signatures associated with the origin of neutrino masses at present and future runs of the LHC.

\section*{Acknowledgments}

TG would like to thank Alexandre Alves, Kuver Sinha and Antonio Oliveira for useful discussions during the early stages of this project. AS thanks Anusandhan National Research Foundation (ANRF) for providing the necessary financial support through the SERB-NPDF grant
(Ref. No.: PDF/2023/002572). AS and TG acknowledges the support from the Department of Atomic Energy (DAE), India, for the Regional Centre for Accelerator-based Particle Physics (RECAPP), Harish-Chandra Research Institute. AD thanks the Carl Trygger Foundation for the support under the project CTS 23:2930. B.M. thanks the Department of Atomic Energy, Government of India,
for support in the form of a Raja Ramanna Chair position.
 

\bibliographystyle{JHEP}
\bibliography{Compressed.bib}

\providecommand{\href}[2]{#2}\begingroup\raggedright\begin{thebibliography}{10}

\bibitem{Konetschny:1977bn}
W.~Konetschny and W.~Kummer, \emph{{Nonconservation of Total Lepton Number with
  Scalar Bosons}},
  \href{https://doi.org/10.1016/0370-2693(77)90407-5}{\emph{Phys. Lett. B}
  {\bfseries 70} (1977) 433}.

\bibitem{Cheng:1980qt}
T.P.~Cheng and L.-F.~Li, \emph{{Neutrino Masses, Mixings and Oscillations in
  SU(2) x U(1) Models of Electroweak Interactions}},
  \href{https://doi.org/10.1103/PhysRevD.22.2860}{\emph{Phys. Rev. D}
  {\bfseries 22} (1980) 2860}.

\bibitem{Lazarides:1980nt}
G.~Lazarides, Q.~Shafi and C.~Wetterich, \emph{{Proton Lifetime and Fermion
  Masses in an SO(10) Model}},
  \href{https://doi.org/10.1016/0550-3213(81)90354-0}{\emph{Nucl. Phys. B}
  {\bfseries 181} (1981) 287}.

\bibitem{Schechter:1980gr}
J.~Schechter and J.W.F.~Valle, \emph{{Neutrino Masses in SU(2) x U(1)
  Theories}}, \href{https://doi.org/10.1103/PhysRevD.22.2227}{\emph{Phys. Rev.
  D} {\bfseries 22} (1980) 2227}.

\bibitem{Mohapatra:1980yp}
R.N.~Mohapatra and G.~Senjanovic, \emph{{Neutrino Masses and Mixings in Gauge
  Models with Spontaneous Parity Violation}},
  \href{https://doi.org/10.1103/PhysRevD.23.165}{\emph{Phys. Rev. D} {\bfseries
  23} (1981) 165}.

\bibitem{Magg:1980ut}
M.~Magg and C.~Wetterich, \emph{{Neutrino Mass Problem and Gauge Hierarchy}},
  \href{https://doi.org/10.1016/0370-2693(80)90825-4}{\emph{Phys. Lett. B}
  {\bfseries 94} (1980) 61}.

\bibitem{ATLAS:2012yve}
{\scshape ATLAS} collaboration, \emph{{Observation of a new particle in the
  search for the Standard Model Higgs boson with the ATLAS detector at the
  LHC}}, \href{https://doi.org/10.1016/j.physletb.2012.08.020}{\emph{Phys.
  Lett. B} {\bfseries 716} (2012) 1}
  [\href{https://arxiv.org/abs/1207.7214}{{\ttfamily 1207.7214}}].

\bibitem{CMS:2012qbp}
{\scshape CMS} collaboration, \emph{{Observation of a New Boson at a Mass of
  125 GeV with the CMS Experiment at the LHC}},
  \href{https://doi.org/10.1016/j.physletb.2012.08.021}{\emph{Phys. Lett. B}
  {\bfseries 716} (2012) 30} [\href{https://arxiv.org/abs/1207.7235}{{\ttfamily
  1207.7235}}].

\bibitem{ATLAS:2022vkf}
{\scshape ATLAS} collaboration, \emph{{A detailed map of Higgs boson
  interactions by the ATLAS experiment ten years after the discovery}},
  \href{https://doi.org/10.1038/s41586-022-04893-w}{\emph{Nature} {\bfseries
  607} (2022) 52} [\href{https://arxiv.org/abs/2207.00092}{{\ttfamily
  2207.00092}}].

\bibitem{CMS:2022dwd}
{\scshape CMS} collaboration, \emph{{A portrait of the Higgs boson by the CMS
  experiment ten years after the discovery.}},
  \href{https://doi.org/10.1038/s41586-022-04892-x}{\emph{Nature} {\bfseries
  607} (2022) 60} [\href{https://arxiv.org/abs/2207.00043}{{\ttfamily
  2207.00043}}].

\bibitem{ATLAS:2012hi}
{\scshape ATLAS} collaboration, \emph{{Search for doubly-charged Higgs bosons
  in like-sign dilepton final states at $\sqrt{s}=7$ TeV with the ATLAS
  detector}}, \href{https://doi.org/10.1140/epjc/s10052-012-2244-2}{\emph{Eur.
  Phys. J. C} {\bfseries 72} (2012) 2244}
  [\href{https://arxiv.org/abs/1210.5070}{{\ttfamily 1210.5070}}].

\bibitem{CMS:2012dun}
{\scshape CMS} collaboration, \emph{{A Search for a Doubly-Charged Higgs Boson
  in $pp$ Collisions at $\sqrt{s}=7$ TeV}},
  \href{https://doi.org/10.1140/epjc/s10052-012-2189-5}{\emph{Eur. Phys. J. C}
  {\bfseries 72} (2012) 2189}
  [\href{https://arxiv.org/abs/1207.2666}{{\ttfamily 1207.2666}}].

\bibitem{ATLAS:2014kca}
{\scshape ATLAS} collaboration, \emph{{Search for anomalous production of
  prompt same-sign lepton pairs and pair-produced doubly charged Higgs bosons
  with $ \sqrt{s}=8 $ TeV $pp$ collisions using the ATLAS detector}},
  \href{https://doi.org/10.1007/JHEP03(2015)041}{\emph{JHEP} {\bfseries 03}
  (2015) 041} [\href{https://arxiv.org/abs/1412.0237}{{\ttfamily 1412.0237}}].

\bibitem{CMS:2014mra}
{\scshape CMS} collaboration, \emph{{Study of vector boson scattering and
  search for new physics in events with two same-sign leptons and two jets}},
  \href{https://doi.org/10.1103/PhysRevLett.114.051801}{\emph{Phys. Rev. Lett.}
  {\bfseries 114} (2015) 051801}
  [\href{https://arxiv.org/abs/1410.6315}{{\ttfamily 1410.6315}}].

\bibitem{CMS:2016cpz}
{\scshape CMS} collaboration, \emph{{Search for a doubly-charged Higgs boson
  with $\sqrt{s}=8~\mathrm{TeV}$ $pp$ collisions at the CMS experiment}}, .

\bibitem{CMS:2017pet}
{\scshape CMS} collaboration, \emph{{A search for doubly-charged Higgs boson
  production in three and four lepton final states at
  $\sqrt{s}=13~\mathrm{TeV}$}}, .

\bibitem{ATLAS:2017xqs}
{\scshape ATLAS} collaboration, \emph{{Search for doubly charged Higgs boson
  production in multi-lepton final states with the ATLAS detector using
  proton{\textendash}proton collisions at $\sqrt{s}=13\,\text {TeV}$}},
  \href{https://doi.org/10.1140/epjc/s10052-018-5661-z}{\emph{Eur. Phys. J. C}
  {\bfseries 78} (2018) 199}
  [\href{https://arxiv.org/abs/1710.09748}{{\ttfamily 1710.09748}}].

\bibitem{CMS:2017fhs}
{\scshape CMS} collaboration, \emph{{Observation of electroweak production of
  same-sign W boson pairs in the two jet and two same-sign lepton final state
  in proton-proton collisions at $\sqrt{s} = $ 13 TeV}},
  \href{https://doi.org/10.1103/PhysRevLett.120.081801}{\emph{Phys. Rev. Lett.}
  {\bfseries 120} (2018) 081801}
  [\href{https://arxiv.org/abs/1709.05822}{{\ttfamily 1709.05822}}].

\bibitem{ATLAS:2018ceg}
{\scshape ATLAS} collaboration, \emph{{Search for doubly charged scalar bosons
  decaying into same-sign $W$ boson pairs with the ATLAS detector}},
  \href{https://doi.org/10.1140/epjc/s10052-018-6500-y}{\emph{Eur. Phys. J. C}
  {\bfseries 79} (2019) 58} [\href{https://arxiv.org/abs/1808.01899}{{\ttfamily
  1808.01899}}].

\bibitem{ATLAS:2021jol}
{\scshape ATLAS} collaboration, \emph{{Search for doubly and singly charged
  Higgs bosons decaying into vector bosons in multi-lepton final states with
  the ATLAS detector using proton-proton collisions at $ \sqrt{\mathrm{s}} $ =
  13 TeV}}, \href{https://doi.org/10.1007/JHEP06(2021)146}{\emph{JHEP}
  {\bfseries 06} (2021) 146}
  [\href{https://arxiv.org/abs/2101.11961}{{\ttfamily 2101.11961}}].

\bibitem{ATLAS:2022pbd}
{\scshape ATLAS} collaboration, \emph{{Search for doubly charged Higgs boson
  production in multi-lepton final states using 139~fb$^{-1}$ of
  proton{\textendash}proton collisions at $\sqrt{s}$ = 13~TeV with the ATLAS
  detector}}, \href{https://doi.org/10.1140/epjc/s10052-023-11578-9}{\emph{Eur.
  Phys. J. C} {\bfseries 83} (2023) 605}
  [\href{https://arxiv.org/abs/2211.07505}{{\ttfamily 2211.07505}}].

\bibitem{Antusch:2018svb}
S.~Antusch, O.~Fischer, A.~Hammad and C.~Scherb, \emph{{Low scale type II
  seesaw: Present constraints and prospects for displaced vertex searches}},
  \href{https://doi.org/10.1007/JHEP02(2019)157}{\emph{JHEP} {\bfseries 02}
  (2019) 157} [\href{https://arxiv.org/abs/1811.03476}{{\ttfamily
  1811.03476}}].

\bibitem{deMelo:2019asm}
T.B.~de~Melo, F.S.~Queiroz and Y.~Villamizar, \emph{{Doubly Charged Scalar at
  the High-Luminosity and High-Energy LHC}},
  \href{https://doi.org/10.1142/S0217751X19501574}{\emph{Int. J. Mod. Phys. A}
  {\bfseries 34} (2019) 1950157}
  [\href{https://arxiv.org/abs/1909.07429}{{\ttfamily 1909.07429}}].

\bibitem{Primulando:2019evb}
R.~Primulando, J.~Julio and P.~Uttayarat, \emph{{Scalar phenomenology in
  type-II seesaw model}},
  \href{https://doi.org/10.1007/JHEP08(2019)024}{\emph{JHEP} {\bfseries 08}
  (2019) 024} [\href{https://arxiv.org/abs/1903.02493}{{\ttfamily
  1903.02493}}].

\bibitem{Chun:2019hce}
E.J.~Chun, S.~Khan, S.~Mandal, M.~Mitra and S.~Shil, \emph{{Same-sign
  tetralepton signature at the Large Hadron Collider and a future $pp$
  collider}}, \href{https://doi.org/10.1103/PhysRevD.101.075008}{\emph{Phys.
  Rev. D} {\bfseries 101} (2020) 075008}
  [\href{https://arxiv.org/abs/1911.00971}{{\ttfamily 1911.00971}}].

\bibitem{Cai:2017mow}
Y.~Cai, T.~Han, T.~Li and R.~Ruiz, \emph{{Lepton Number Violation: Seesaw
  Models and Their Collider Tests}},
  \href{https://doi.org/10.3389/fphy.2018.00040}{\emph{Front. in Phys.}
  {\bfseries 6} (2018) 40} [\href{https://arxiv.org/abs/1711.02180}{{\ttfamily
  1711.02180}}].

\bibitem{Deppisch:2015qwa}
F.F.~Deppisch, P.S.~Bhupal~Dev and A.~Pilaftsis, \emph{{Neutrinos and Collider
  Physics}}, \href{https://doi.org/10.1088/1367-2630/17/7/075019}{\emph{New J.
  Phys.} {\bfseries 17} (2015) 075019}
  [\href{https://arxiv.org/abs/1502.06541}{{\ttfamily 1502.06541}}].

\bibitem{Ashanujjaman:2021txz}
S.~Ashanujjaman and K.~Ghosh, \emph{{Revisiting type-II see-saw: present limits
  and future prospects at LHC}},
  \href{https://doi.org/10.1007/JHEP03(2022)195}{\emph{JHEP} {\bfseries 03}
  (2022) 195} [\href{https://arxiv.org/abs/2108.10952}{{\ttfamily
  2108.10952}}].

\bibitem{Ashanujjaman:2023tlj}
S.~Ashanujjaman and S.P.~Maharathy, \emph{{Probing compressed mass spectra in
  the type-II seesaw model at the LHC}},
  \href{https://doi.org/10.1103/PhysRevD.107.115026}{\emph{Phys. Rev. D}
  {\bfseries 107} (2023) 115026}
  [\href{https://arxiv.org/abs/2305.06889}{{\ttfamily 2305.06889}}].

\bibitem{ATLAS:2016wtr}
{\scshape ATLAS} collaboration, \emph{{Performance of the ATLAS Trigger System
  in 2015}}, \href{https://doi.org/10.1140/epjc/s10052-017-4852-3}{\emph{Eur.
  Phys. J. C} {\bfseries 77} (2017) 317}
  [\href{https://arxiv.org/abs/1611.09661}{{\ttfamily 1611.09661}}].

\bibitem{ATLAS:2021moa}
{\scshape ATLAS} collaboration, \emph{{Search for
  chargino{\textendash}neutralino pair production in final states with three
  leptons and missing transverse momentum in $\sqrt{s} = 13$~TeV pp collisions
  with the ATLAS detector}},
  \href{https://doi.org/10.1140/epjc/s10052-021-09749-7}{\emph{Eur. Phys. J. C}
  {\bfseries 81} (2021) 1118}
  [\href{https://arxiv.org/abs/2106.01676}{{\ttfamily 2106.01676}}].

\bibitem{BhupalDev:2013xol}
P.S.~Bhupal~Dev, D.K.~Ghosh, N.~Okada and I.~Saha, \emph{{125 GeV Higgs Boson
  and the Type-II Seesaw Model}},
  \href{https://doi.org/10.1007/JHEP03(2013)150}{\emph{JHEP} {\bfseries 03}
  (2013) 150} [\href{https://arxiv.org/abs/1301.3453}{{\ttfamily 1301.3453}}].

\bibitem{Dev:2017ouk}
P.S.B.~Dev, C.M.~Vila and W.~Rodejohann, \emph{{Naturalness in testable type II
  seesaw scenarios}},
  \href{https://doi.org/10.1016/j.nuclphysb.2017.06.007}{\emph{Nucl. Phys. B}
  {\bfseries 921} (2017) 436}
  [\href{https://arxiv.org/abs/1703.00828}{{\ttfamily 1703.00828}}].

\bibitem{ParticleDataGroup:2020ssz}
{\scshape Particle Data Group} collaboration, \emph{{Review of Particle
  Physics}}, \href{https://doi.org/10.1093/ptep/ptaa104}{\emph{PTEP} {\bfseries
  2020} (2020) 083C01}.

\bibitem{Esteban:2020cvm}
I.~Esteban, M.C.~Gonzalez-Garcia, M.~Maltoni, T.~Schwetz and A.~Zhou,
  \emph{{The fate of hints: updated global analysis of three-flavor neutrino
  oscillations}}, \href{https://doi.org/10.1007/JHEP09(2020)178}{\emph{JHEP}
  {\bfseries 09} (2020) 178}
  [\href{https://arxiv.org/abs/2007.14792}{{\ttfamily 2007.14792}}].

\bibitem{Das:2016bir}
D.~Das and A.~Santamaria, \emph{{Updated scalar sector constraints in the Higgs
  triplet model}},
  \href{https://doi.org/10.1103/PhysRevD.94.015015}{\emph{Phys. Rev. D}
  {\bfseries 94} (2016) 015015}
  [\href{https://arxiv.org/abs/1604.08099}{{\ttfamily 1604.08099}}].

\bibitem{Melfo:2011nx}
A.~Melfo, M.~Nemevsek, F.~Nesti, G.~Senjanovic and Y.~Zhang, \emph{{Type II
  Seesaw at LHC: The Roadmap}},
  \href{https://doi.org/10.1103/PhysRevD.85.055018}{\emph{Phys. Rev. D}
  {\bfseries 85} (2012) 055018}
  [\href{https://arxiv.org/abs/1108.4416}{{\ttfamily 1108.4416}}].

\bibitem{Chun:2012jw}
E.J.~Chun, H.M.~Lee and P.~Sharma, \emph{{Vacuum Stability, Perturbativity,
  EWPD and Higgs-to-diphoton rate in Type II Seesaw Models}},
  \href{https://doi.org/10.1007/JHEP11(2012)106}{\emph{JHEP} {\bfseries 11}
  (2012) 106} [\href{https://arxiv.org/abs/1209.1303}{{\ttfamily 1209.1303}}].

\bibitem{Aoki:2012jj}
M.~Aoki, S.~Kanemura, M.~Kikuchi and K.~Yagyu, \emph{{Radiative corrections to
  the Higgs boson couplings in the triplet model}},
  \href{https://doi.org/10.1103/PhysRevD.87.015012}{\emph{Phys. Rev. D}
  {\bfseries 87} (2013) 015012}
  [\href{https://arxiv.org/abs/1211.6029}{{\ttfamily 1211.6029}}].

\bibitem{Dey:2020tfq}
A.~Dey, J.~Lahiri and B.~Mukhopadhyaya, \emph{{LHC signals of triplet scalars
  as dark matter portal: cut-based approach and improvement with gradient
  boosting and neural networks}},
  \href{https://doi.org/10.1007/JHEP06(2020)126}{\emph{JHEP} {\bfseries 06}
  (2020) 126} [\href{https://arxiv.org/abs/2001.09349}{{\ttfamily
  2001.09349}}].

\bibitem{Baak:2014ora}
{\scshape Gfitter Group} collaboration, \emph{{The global electroweak fit at
  NNLO and prospects for the LHC and ILC}},
  \href{https://doi.org/10.1140/epjc/s10052-014-3046-5}{\emph{Eur. Phys. J. C}
  {\bfseries 74} (2014) 3046}
  [\href{https://arxiv.org/abs/1407.3792}{{\ttfamily 1407.3792}}].

\bibitem{Dinh:2012bp}
D.N.~Dinh, A.~Ibarra, E.~Molinaro and S.T.~Petcov, \emph{{The $\mu - e$
  Conversion in Nuclei, $\mu \to e \gamma, \mu \to 3e$ Decays and TeV Scale
  See-Saw Scenarios of Neutrino Mass Generation}},
  \href{https://doi.org/10.1007/JHEP08(2012)125}{\emph{JHEP} {\bfseries 08}
  (2012) 125} [\href{https://arxiv.org/abs/1205.4671}{{\ttfamily 1205.4671}}].

\bibitem{Akeroyd:2009nu}
A.G.~Akeroyd, M.~Aoki and H.~Sugiyama, \emph{{Lepton Flavour Violating Decays
  tau ---{\ensuremath{>}} anti-l ll and mu ---{\ensuremath{>}} e gamma in the
  Higgs Triplet Model}},
  \href{https://doi.org/10.1103/PhysRevD.79.113010}{\emph{Phys. Rev. D}
  {\bfseries 79} (2009) 113010}
  [\href{https://arxiv.org/abs/0904.3640}{{\ttfamily 0904.3640}}].

\bibitem{Kakizaki:2003jk}
M.~Kakizaki, Y.~Ogura and F.~Shima, \emph{{Lepton flavor violation in the
  triplet Higgs model}},
  \href{https://doi.org/10.1016/S0370-2693(03)00833-5}{\emph{Phys. Lett. B}
  {\bfseries 566} (2003) 210}
  [\href{https://arxiv.org/abs/hep-ph/0304254}{{\ttfamily hep-ph/0304254}}].

\bibitem{MEG:2016leq}
{\scshape MEG} collaboration, \emph{{Search for the lepton flavour violating
  decay $\mu ^+ \rightarrow \mathrm {e}^+ \gamma $ with the full dataset of the
  MEG experiment}},
  \href{https://doi.org/10.1140/epjc/s10052-016-4271-x}{\emph{Eur. Phys. J. C}
  {\bfseries 76} (2016) 434}
  [\href{https://arxiv.org/abs/1605.05081}{{\ttfamily 1605.05081}}].

\bibitem{SINDRUM:1987nra}
{\scshape SINDRUM} collaboration, \emph{{Search for the Decay $\mu^+ \to e^+
  e^+ e^-$}}, \href{https://doi.org/10.1016/0550-3213(88)90462-2}{\emph{Nucl.
  Phys. B} {\bfseries 299} (1988) 1}.

\bibitem{Ashanujjaman:2025scr}
S.~Ashanujjaman, P.S.B.~Dev, J.~Huang and S.~Zhou, \emph{{Precision Higgs Probe
  of Type-II Seesaw}},  \href{https://arxiv.org/abs/2512.07532}{{\ttfamily
  2512.07532}}.

\bibitem{Schwaller:2013baa}
P.~Schwaller and J.~Zurita, \emph{{Compressed electroweakino spectra at the
  LHC}}, \href{https://doi.org/10.1007/JHEP03(2014)060}{\emph{JHEP} {\bfseries
  03} (2014) 060} [\href{https://arxiv.org/abs/1312.7350}{{\ttfamily
  1312.7350}}].

\bibitem{Baer:2014kya}
H.~Baer, A.~Mustafayev and X.~Tata, \emph{{Monojet plus soft dilepton signal
  from light higgsino pair production at LHC14}},
  \href{https://doi.org/10.1103/PhysRevD.90.115007}{\emph{Phys. Rev. D}
  {\bfseries 90} (2014) 115007}
  [\href{https://arxiv.org/abs/1409.7058}{{\ttfamily 1409.7058}}].

\bibitem{Dutta:2017nqv}
B.~Dutta, K.~Fantahun, A.~Fernando, T.~Ghosh, J.~Kumar, P.~Sandick et~al.,
  \emph{{Probing Squeezed Bino-Slepton Spectra with the Large Hadron
  Collider}}, \href{https://doi.org/10.1103/PhysRevD.96.075037}{\emph{Phys.
  Rev. D} {\bfseries 96} (2017) 075037}
  [\href{https://arxiv.org/abs/1706.05339}{{\ttfamily 1706.05339}}].

\bibitem{Alloul:2013bka}
A.~Alloul, N.D.~Christensen, C.~Degrande, C.~Duhr and B.~Fuks, \emph{{FeynRules
  2.0 - A complete toolbox for tree-level phenomenology}},
  \href{https://doi.org/10.1016/j.cpc.2014.04.012}{\emph{Comput. Phys. Commun.}
  {\bfseries 185} (2014) 2250}
  [\href{https://arxiv.org/abs/1310.1921}{{\ttfamily 1310.1921}}].

\bibitem{Alwall:2014hca}
J.~Alwall, R.~Frederix, S.~Frixione, V.~Hirschi, F.~Maltoni, O.~Mattelaer
  et~al., \emph{{The automated computation of tree-level and next-to-leading
  order differential cross sections, and their matching to parton shower
  simulations}}, \href{https://doi.org/10.1007/JHEP07(2014)079}{\emph{JHEP}
  {\bfseries 07} (2014) 079} [\href{https://arxiv.org/abs/1405.0301}{{\ttfamily
  1405.0301}}].

\bibitem{Ball:2012cx}
R.D.~Ball et~al., \emph{{Parton distributions with LHC data}},
  \href{https://doi.org/10.1016/j.nuclphysb.2012.10.003}{\emph{Nucl. Phys. B}
  {\bfseries 867} (2013) 244}
  [\href{https://arxiv.org/abs/1207.1303}{{\ttfamily 1207.1303}}].

\bibitem{Bierlich:2022pfr}
C.~Bierlich et~al., \emph{{A comprehensive guide to the physics and usage of
  PYTHIA 8.3}},
  \href{https://doi.org/10.21468/SciPostPhysCodeb.8}{\emph{SciPost Phys.
  Codeb.} {\bfseries 2022} (2022) 8}
  [\href{https://arxiv.org/abs/2203.11601}{{\ttfamily 2203.11601}}].

\bibitem{deFavereau:2013fsa}
{\scshape DELPHES 3} collaboration, \emph{{DELPHES 3, A modular framework for
  fast simulation of a generic collider experiment}},
  \href{https://doi.org/10.1007/JHEP02(2014)057}{\emph{JHEP} {\bfseries 02}
  (2014) 057} [\href{https://arxiv.org/abs/1307.6346}{{\ttfamily 1307.6346}}].

\bibitem{Mangano:2006rw}
M.L.~Mangano, M.~Moretti, F.~Piccinini and M.~Treccani, \emph{{Matching matrix
  elements and shower evolution for top-quark production in hadronic
  collisions}},
  \href{https://doi.org/10.1088/1126-6708/2007/01/013}{\emph{JHEP} {\bfseries
  01} (2007) 013} [\href{https://arxiv.org/abs/hep-ph/0611129}{{\ttfamily
  hep-ph/0611129}}].

\bibitem{Cacciari:2011ma}
M.~Cacciari, G.P.~Salam and G.~Soyez, \emph{{FastJet User Manual}},
  \href{https://doi.org/10.1140/epjc/s10052-012-1896-2}{\emph{Eur. Phys. J. C}
  {\bfseries 72} (2012) 1896}
  [\href{https://arxiv.org/abs/1111.6097}{{\ttfamily 1111.6097}}].

\end{thebibliography}\endgroup
\end{document}